\documentclass[apj]{emulateapj}
\pdfoutput=1
\usepackage{graphics}
\usepackage{hyperref}
\usepackage{enumerate}
\usepackage{color}

\usepackage{amsmath}
\usepackage{threeparttable}

\newcommand{\revone}[1]{{ #1} }

\shortauthors{Nandez et al.}
\shorttitle{V1309 Sco -- Understanding a merger}
\begin{document}
\title{V1309 SCO -- Understanding a merger}

\author{J.L.A. Nandez}
\email{avendaon@ualberta.ca}
\author{N. Ivanova}
\affil{Department of Physics, University of Alberta, Edmonton, AB, T6G 2E7, Canada}
\author{J.C. Lombardi, Jr.}
\affil{Department of Physics, Allegheny College, Meadville, PA 16335, USA}


\begin{abstract}{
One of the two outcomes of a common envelope event is a merger of the two stars. To date, the best known case of a binary merger is the V1309~Sco outburst, where the orbital period was known and observed to decay up to the outburst. Using the hydrodynamical code {\tt StarSmasher}, we study in detail which characteristics of the progenitor binary affect the outburst and produce the best match with observations. We have developed a set of tools in order to quantify any common envelope event. We consider binaries consisting of a 1.52$M_\odot$ giant and a 0.16$M_\odot$ companion with $P_{\rm orb}\sim1.4$ days, varying the nature of the companion and its synchronization. We show that all considered progenitor binaries evolve towards the merger primarily due to the Darwin instability. The merger is accompanied by mass ejection that proceeds in several separate mass outbursts and takes away a few percent of the donor mass. This very small mass, nonetheless, is important as it is not only sufficient to explain the observed light-curve, but it also carries away up to 1/3 of the both initial total angular momentum and initial orbital energy. We find that all synchronized systems experience $L_2$ mass loss that operates during just a few days prior to merger and produces ring-shaped ejecta. The formed star is always a strongly heated radiative star that differentially rotates. We conclude that the case of a synchronized binary with a main-sequence companion gives the best match with observations of V1309~Sco.
}
\end{abstract}



\section{Introduction}
\label{sec:introduction}

More than half of all stars are in binaries or systems of even higher multiplicity (triplets, etc.) --
e.g., as many as two-thirds of G stars are in multiple systems \citep{1991A&A...248..485D}.  
The binary fraction increases with the spectral class, and for massive stars it is so large that
more than 70\% of them are expected to exchange mass with a companion \citep{2012Sci...337..444S}.
The fate of a binary is decided by how stable or unstable this mass transfer is.
If the companion is not able to accept all the transferred mass, 
then the two stars start to share their outer layers --
forming a so-called ``common envelope'' (CE). 
The outcome of the CE depends on how much orbital energy is deposited in the envelope. 
If the binary deposits enough orbital energy, the envelope could be ejected and a new binary 
composed of the companion and the core of the donor will be left in a tight orbit \citep{Pa76,1984ApJ...277..355W}. 
However, if the binary does not deposit enough energy, 
it will instead merge to form a single star \citep[for more details and for the most recent review of the CE event, see][]{2013A&ARv..21...59I}.

A new class of transients was recently identified.
With a total energy output anywhere in the range of $10^{45}-10^{47}$ ergs,  their peak luminosities were just below that of Type Ia SNe while still above that of novae \citep[see e.g.][]{2003Natur.422..405B,2007Natur.447..458K,2011ApJ...737...17B}.
The spectra of these mysterious transients are predominantly red, completely unlike novae, and they are known under several alternate names -- 
Luminous Red Novae (LRNe),  Intermediate Luminosity Red Transients (ILRTs), Intermediate Luminosity Optical Transients (ILOTs), V838 Mon-like events and supernova impostors.
In this work, we pay specific attention to those of red transients which are usually labeled as LRNe, with the most important examples being 
V838 Mon \citep{2002MNRAS.336L..43K},  V1309~Sco \citep{2011A&A...528A.114T}, M85~OT2006-1 \citep{2007ApJ...659.1536R} and
M31~RV \citep{2004A&A...418..869B}. 
We note that the LRN class is likely distinct in its nature from another class of red transients, known as supernova impostors.  Their progenitors 
have been observationally identified as dusty modestly-massive stars  – as for example in the cases of SN 2008S and NGC 300 OT \citep{2008ApJ...681L...9P,2009ApJ...699.1850B,2009ApJ...697L..49S}. 

Several hypotheses have been proposed to explain the nature of outbursts in different LRNe. 
The most common model is a merger,
either of two stars or a star and a planet, with such merger-burst models being  
first proposed to explain V838 Mon type events 
\citep{2003ApJ...582L.105S, 2006A&A...451..223T}.
Later, \cite{2012ApJ...746..100S} also suggested that the outbursts could be powered by mass accretion 
onto a main-sequence star from an asymptotic giant branch star. 

There were however initially some problems with justifying the physics behind the merger model. 
For example, \cite{2004A&A...418..869B} had noticed that the M31-RV, V838 Mon, and V4332 Sgr outbursts were strongly 
homological and could not be explained by merger-powered outbursts
that showed too much dependency on metallicity, mass, and ages.
A simplistic estimate of the energy that could be available from a merger 
fell short by a factor of a few from the the energy that was radiated away from M85 OT2006-1 \citep{Of08}.
\cite{2013Sci...339..433I} provided a further comparison between the theoretical expectations for a merger-burst model  
and the observations and found a number of other inconsistencies.
In particular, compared to theoretical expectations, the observations implied too large of an increase in radius and luminosity, as well as too long in duration for the outburst and
plateau phases.  In addition, the observed velocities and extremely rapid luminosity decline were difficult to explain.

The Rosetta Stone was the V1309 Sco outburst, which was observed before, during and after its outburst. 
The key was that the observations showed that the object was a contact binary prior to the outburst and a single object afterwards \citep{2011A&A...528A.114T}, undoubtedly indicating a merger.
Based on these observations, \cite{2013Sci...339..433I} suggested that the outburst in V1309 Sco, as well as in similar LRNe, 
is controlled by the recombination of the material that is ejected during the CE event. 
This model helps to explain the homology of the class including the plateau phase,
the range of the total energy radiated away, the differences between the velocities 
derived from a spectra and from apparent radius expansion as well as the observed colors.
The proposed link between the observations and the theoretically predicted light-curve relies 
on how much of the material is ejected and how much kinetic energy that material carries away.

In this paper we present numerical simulations of the merger of the V1309 Sco binary, describing in full detail 
the models that were used to predict the light-curve of V1309 Sco in \cite{2013Sci...339..433I}. 
We depict our numerical methods, initial models, and assumptions in \S2. 
To make useful predictions that would allow the linking of theory and observations, 
we pay special attention in \S3 to how to quantify key quantities in the merger event -- what is  bound and unbound material, 
the entropy  and temperature of the ejecta -- as well as how to determine when the physical merger takes place.
We discuss how the pre-merger binary evolves to the CE events in \S4. 
This includes how the initial conditions (such as binary's synchronization and the nature of the companion) could affect its orbital evolution. 
We also discuss whether it is possible to  match the period decay observed in V1309 Sco.
The details of the mass exchange and the mass loss prior to the merger are also described in \S4, 
while the mass ejection throughout the whole process is discussed in \S5. 
Further discussion on how to classify the unbound material and its properties at the end of the simulation 
are also given in \S5. 
In \S6 we talk about the symmetry of the merger product, how to get 1D profiles from the 3D SPH code,
and what the entropy and rotation profiles the merger product are.

\begin{table*}[htp!]
\caption{Initial conditions for the performed simulations in this work.} \label{tab:tablesol}
\begin{center}
\begin{tabular}{l l l l l l l l l l r  r c }
\hline
 Model  &$R_{\rm RG}^{\rm ev} $ & $R_{\rm RG}$ & $R_*$ & $R_V$ &$a $  & $P_{\rm orb} $ & $f_{\rm sync}$ &$f_{\rm RLOF}$&$f^*_{\rm RLOF}$ &$N_{\rm 1}$ &$N_{\rm 2}$& Companion Star \\
\hline
 ps334 & 3.40 & 3.34 & 3.56 & 3.26 &6.32 & 1.42 & 0.915 & 0.90 & 0.98 & 50161 & 1 &Degenerate  \\
 mn351 & 3.40 & 3.51 & 3.70 & 3.52 &6.32 & 1.42 & 0.000 & 0.97 & 1.02 & 99955& 19938 &Main-Sequence \\
 pn351 & 3.40 & 3.51 & 3.70 & 3.52 &6.32 & 1.42 & 0.000 & 0.97 & 1.02& 99955 & 1 &Degenerate \\
 ps351 & 3.40 & 3.51 & 3.70 & 3.39 &6.32 & 1.42 & 1.000 & 0.93 & 1.02 & 169831 & 1 & Degenerate  \\
 ms376 & 3.65 & 3.76 & 3.96 & 3.63 &6.55 & 1.50 & 1.000 & 0.97 & 1.05 & 99955 & 1974 &Main-Sequence \\
 ps376 & 3.65 & 3.76 & 3.96 & 3.63 &6.55 & 1.50 & 1.000 & 0.97 & 1.05 & 99955 & 1 &Degenerate  \\
 ms372 & 3.70 & 3.72 & 3.98 & 3.59 &6.40 & 1.45 & 0.937 & 0.98 & 1.08& 50161& 1974 & Main-Sequence \\
 ps379 & 3.73 & 3.79 & 4.00 & 3.68 &6.32 & 1.42 & 0.854 & 1.01 & 1.10& 99955& 1 & Degenerate\\
 pn344 & 3.66 & 3.44 & 3.63 & 3.46 &6.38 & 1.44 & 0.000 & 0.94 & 0.99& 99955 & 1 & Degenerate\\
 ps375 & 3.66 & 3.75 & 3.95 & 3.63 &6.38 & 1.44 & 1.000 & 0.99 & 1.08& 99955 & 1 & Degenerate\\
 mn344 & 3.66 & 3.44 & 3.63 & 3.46 &6.38 & 1.44 & 0.000 & 0.94 & 0.99& 99955 & 4944 & Main-Sequence\\
 ms375 & 3.66 & 3.75 & 3.95 & 3.63 &6.38 & 1.44 & 1.000 & 0.99 & 1.08&99955 & 4944 & Main-Sequence\\
 pn319 & 3.39 & 3.19 & 3.38 & 3.20 &6.38 & 1.44 & 0.000 & 0.87 & 0.92& 80023 & 1    & Degenerate\\
\hline
\end{tabular}
\begin{tablenotes}
       \item \underline{Name of the model:} \textit{p} stands for a point mass secondary, \textit{m} stands for a main sequence secondary, \textit{n} is for non-synchronized cases, \textit{s} is for synchronized cases, and three digits stand for the value of the relaxed primary radius.
       \item \underline{Radii:} $R_{\rm RG}^{\rm ev}$ is the radius of the donor in the stellar code in $R_\odot$, $R_{\rm RG}$ is the radius of the donor assuming the outermost particle distance in $R_\odot$, $R_*$ is the radius of the donor after adding 1 smoothing lengths for the outermost particle in $R_\odot$, and, $R_V$ is the volume-equivalent radius of the donor.
       \item \underline{Binary initial setup:} $a$ is the orbital separation in $R_\odot$, $P_{\rm orb}$ is the orbital period of an initially relaxed binary in SPH code in days, $f_{\rm sync}$ is the degree of synchronization;  $f_{\rm RLOF}$ is the ratio $R_V$ to the radius of the Roche lobe and characterizes overflow of the donor,  $f^*_{\rm RLOF}$ is the ratio $R_*$ to the radius of the Roche lobe. \\ 
       \item \underline{Resolution:} $N_{\rm 1}$ is the number of particles for the giant stars, $N_{\rm 2}$ is the number of particle for the accreting star.
     \end{tablenotes}
\end{center}
\end{table*}

\section{Modelling the merger: methods and initial conditions}

Observations of V1309 Sco during 2002-2008, presented in \cite{2011A&A...528A.114T}, 
show that the object was a binary with a steadily decaying orbital period $P$ near 1.44 days.   
The binary was argued to be a contact binary with 
the observationally derived effective temperature of $T_{\rm eff}\sim 4500$ K and luminosity of $3.0-8.6\ L_\odot$. 

\cite{2011A&A...531A..18S} used these pre-outburst observations to determine a possible binary 
configuration and the evolutionary states of the progenitor binary companions.
Based on that study, we adopted for our initial conditions a primary star mass $M_1=1.52\ M_\odot$ and a secondary star mass $M_2=0.16\ M_\odot$;
the primary star is an early subgiant while the secondary could be either 
a low-mass main-sequence star or a stripped giant core (essentially, a white dwarf) 
remaining from the previous mass transfer. 

To model the merging binary, we first evolved both companions individually, 
using the stellar evolution code {\tt EV/STARS} 
(\citet{1971MNRAS.151..351E, 1972MNRAS.156..361E}, recently updated \citet{2008A&A...488.1007G}).
The $M_1=1.52\ M_\odot$ primary was evolved until we could match the observations
for temperature, luminosity, and radius (to fulfill the requirement to fill the Roche lobe for the known orbital solution).
When a  lower-mass companion was adopted to be a main sequence star, we evolved it using the same stellar code {\tt EV/STARS}
to the same age as the primary star.
When a lower-mass companion was  a white dwarf -- the case when the secondary could be a stripped core of a red giant and was more massive 
in the beginning -- we did not  model the possible first mass transfer onto current primary, 
as that mass gain that had occurred to the current primary a long time ago, 
during its main sequence and does not significantly affect its current subgiant structure.

At the second step, we used the 1-dimensional stellar structures (obtained from the stellar evolution code) as initial conditions for
3-dimensional (3D) hydrodynamical simulations.
For the 3D simulations, we used the code {\tt StarSmasher}, which is based on
the smoothed particle hydrodynamics (SPH) method; see for example \cite{1992ARA&A..30..543M}. The code was developed 
by \cite{2006ApJ...640..441L} and  the equations of motion have been updated  by \cite{2010MNRAS.402..105G} 
\citep[the most recent version of the code is described in][]{2011ApJ...737...49L}. 

Each star was relaxed in the SPH code individually, by evolving the profile provided by the stellar code to its hydrostatic equilibrium in 3D. 
The relaxed stars then were placed in the inertial frame of a binary.  

In our studies, we performed thirteen merger simulations. The list of all merger  models with the corresponding initial conditions  can be found in 
Table~\ref{tab:tablesol}. Below we describe in detail the reasons for the diversity of the adopted initial conditions. 

\subsection{Nature of the low-mass companion and the number of SPH particles}
In order to find the best scenario to the observations presented by \cite{2011A&A...528A.114T}, 
we compared two possibilities for the nature of the low-mass secondary: a main sequence star
and a stripped core of a red giant (a degenerate companion). 
A degenerate companion is considered as a compact object particle in SPH, 
characterized by its mass and interacting only gravitationally with other SPH particles. 
A main-sequence star was treated as described above, and was generally represented by several thousand SPH particles. 

The number of SPH particles that represent the primary star and the secondary companion need to differ by about an order of magnitude, with more SPH particles representing the primary star. 
This does not correlate directly with the masses of the companions, but rather with their average densities. 
To have comparable smoothing lengths for particles inside a main sequence star and inside a sub-giant,  
the number of SPH particles that describes the secondary must be much smaller than the number for the primary. 
If instead a small companion is represented by a comparable number of particles as the primary, then when the companion is crushed inside the primary star during the merger, 
computational time increases substantially.


In Table~\ref{tab:tablesol} we list the numbers of particles adopted at the start of each simulation: $N_{\rm 1}$ is for the primary star and $N_{\rm 2}$ is for the secondary star.
If $N_{\rm 2}=1$, the companion is modeled as a compact object particle. 

\subsection{The radius of the primary star} 
Mapping a one-dimensional star of radius $R_{\rm RG}^{\rm ev}$ into a 3D star and then relaxing it in a 3D code usually leads to a change of the star's radius
\citep[see also the discussion about the somewhat similar effect for polytropic stars in][]{2002A&A...389..485R}.
Further uncertainty arises from extracting the radius of a 3D star represented by particles instead of on a continuous grid. 
Also, in stellar codes, the stellar radius is by definition the radius of the photosphere, 
which cannot be resolved by our SPH code.
One way to define the stellar radius in 3D is to find the position of the outermost particle.  However, the outermost particle's kernel extends the density to $2h_{\rm out}$ from the location of this particle \citep[see for more details][]{1985A&A...149..135M}, where $h_{\rm out}$ is the smoothing length of the outermost particle. 

Let us consider the mapping and relaxation in more detail.
When a star is first mapped into 3D, the outermost particles will be located at a position $R_{\rm RG}$ that is about 
$2h_{\rm out}$ less than $R_{\rm RG}^{\rm ev}$.
While the relaxation proceeds,  the position of the outermost particle can change, and this change depends on the number of particles, rotation of the star and the method used to relax the star, for example if artificial drag force and/or artificial viscosity is used. Hence $R_{\rm RG}$ at the end of the relaxation can  in some cases be smaller and in other cases larger than $R_{\rm RG}^{\rm ev}$.
In the relaxed model, the density goes to zero at different radii $R$, with $R$ being a function of the polar angle $\theta$ and azimuthal angle $\varphi$.  
The quantity $R_{\rm RG}+2 h_{\rm out}$  is the maximum of $R(\theta, \varphi)$ over all possible $\theta$ and $\varphi$ values.  
Because the density is sometimes zero inside the radius $R_{\rm RG}+2h_{\rm out}$
and always zero outside this radius, the radius $R_{\rm RG}+2h_{\rm out}$ is an overestimate of the average radius.

In Table~\ref{tab:tablesol} we list the primary radii found by several methods. 
$R_{\rm RG}^{\rm ev}$ is the radius as obtained by the one-dimensional stellar evolutionary code,  
$R_{\rm RG}$ is the radius after relaxation in 3D SPH code determined by the outermost particle, and the effective radius $R_*\equiv R_{\rm RG}+h_{\rm out}$.  
In all the cases, the desired radius is within a smoothing length from either $R_{\rm RG}$ or $R_*$.

Arguably the most important radius is the ``volume-equivalent'' radius $R_{\rm V}$. 
In this case, we sum up over all particles $m_i/\rho_i$, where $m_i$ and $\rho_i$ are the mass and density of each particle $i$, to find the total volume $V$ occupied by the particles. Then we solve for radius as $R_{V}= (3 V/{4\pi})^{1/3}$. We find that $R_{V}$ tends to be on the lower boundary of our other radius estimates, very close to $R_{\rm RG}$, and never exceeding $R_*$.
Through this paper, we use $R_{\rm V}$ as the default definition of the primary radius.

\subsection{Orbital separation, orbital period, and the Roche lobe overflow}
The binary orbital separation $a$, the orbital period $P_{\rm orb}$,  and
the ratio of the donor star radius to its Roche lobe radius that
quantifies the Roche lobe overflow (RLOF), $f_{\rm RLOF}=R_{\rm V}/R_{\rm RL}$, are all closely
connected.
With the approximation from \cite{1983ApJ...268..368E}, in our system $R_{\rm RL}=0.574 a$.

While the observed orbital period right before the merger of V1309 Sco was measured to a 
quite good precision, it can not be stated firmly if that period should be 
a true initial period for our merger simulations.
Further, small variance of the pre-merger orbital period  would not affect the outcome qualitatively; however
for our simulations  $f_{\rm RLOF}$ is
very influential on how fast an initial binary would decay into a complete merger.

Adopted initial values for $a$, $P_{\rm orb}$, and $f_{\rm RLOF}$ are listed in Table~\ref{tab:tablesol}.
In the adopted notation, if $f_{\rm RLOF}>1$, the donor star is overflowing, and if $f_{\rm RLOF}<1$, the donor is still confined in its Roche lobe.
We also list  $f^*_{\rm RLOF}=R_*/R_{\rm RL}$. If $f^*_{\rm RLOF} > 1$, a donor may start to loose particles due to their oscillation 
around their positions by a smoothing length, even if $f_{\rm RLOF}<1$.

\subsection{Synchronization}

While stars with convective envelope are believed to be quickly tidally synchronized, it is not fully clear if a relatively fast expanding subgiant 
will remain tidally locked to its ten times less massive companion.
We therefore considered cases with different synchronization, from non-rotating stars to fully synchronized cases. 
In a fully synchronized binary, the angular velocities of both companions are the same as that of the orbit $\Omega_{*,1}=\Omega_{*,2}=\Omega_{\rm orb} = 2\pi /P_{\rm orb}$. 
To quantify the degree of synchronization in each simulation, we introduce $f_{\rm syn} \equiv \Omega_{*}/\Omega_{orb}$. The critical value $f_{\rm syn}=1$ corresponds to a fully synchronized case, 
while $f_{\rm syn}=0$ corresponds to an fully non-synchronized (irrotational) case: see Table~\ref{tab:tablesol}. 
In our simulations, each star is first relaxed with its own spin and only then placed in a binary.

\section{Definitions}

\subsection{Orbital period}

In a system that is not represented by only two mere point masses, 
but by a collection of many particles, the definition of what 
exactly is an orbital period is ambiguous. 
While several approaches can be used, we will discuss and use in this paper two of them.

\begin{figure}[t]
 \begin{center}
   \includegraphics[scale=0.45]{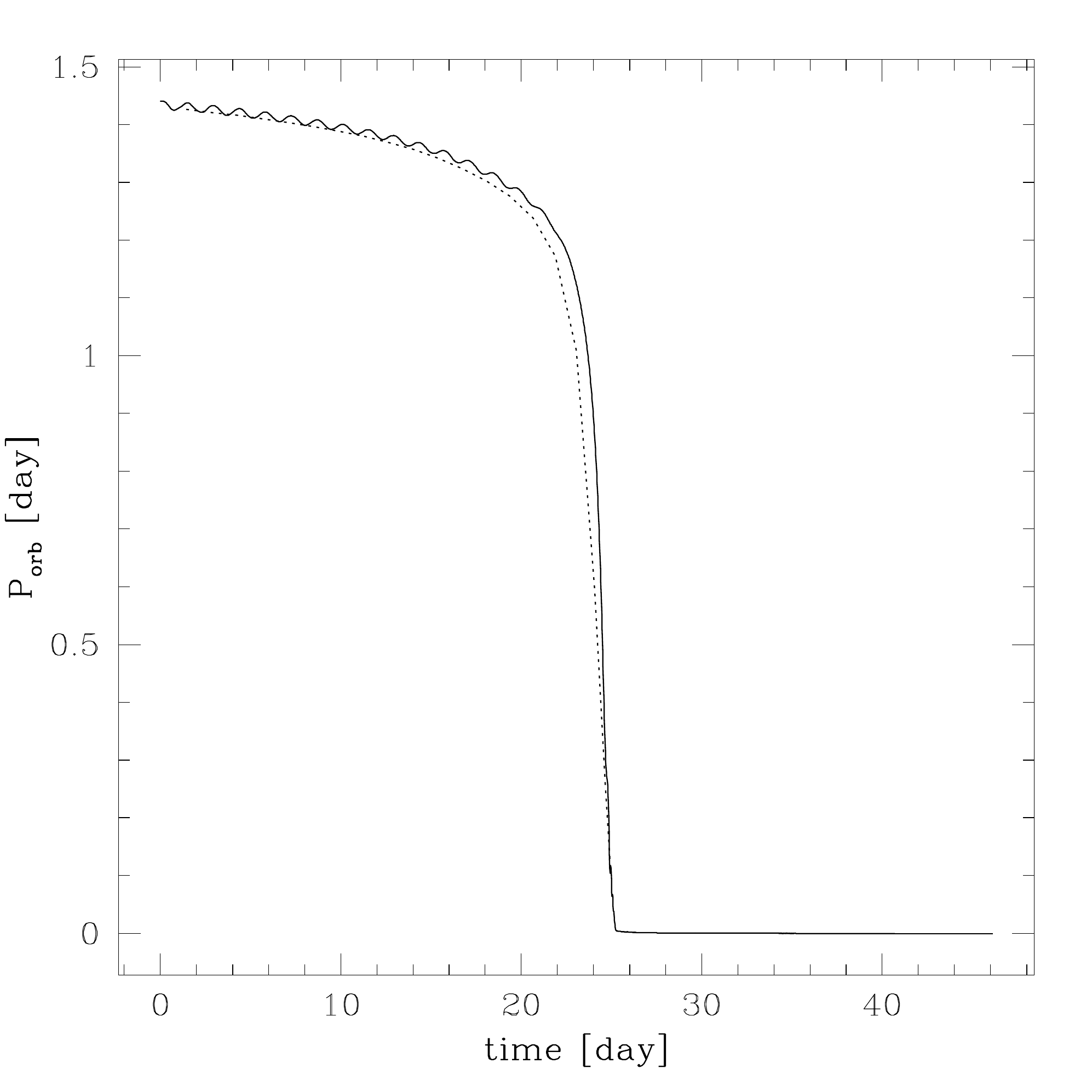}
 \end{center}
\caption{The orbital period for the simulation pn344. 
The instantaneous orbital period is shown with the solid line, and the 
apparent orbital period is shown with the dotted line.}
\label{fig:Cperiod}
\end{figure}

First, we can find the {\it instantaneous} orbital period $P_{\rm orb, inst}$ at each time step. 
This can be done by assuming a Keplerian orbit of two bodies, where the mass
 of each body is the  mass bound to that companion, 
and the separation is given by the locations of the cores of the stars \citep[for more details, see][]{2006ApJ...640..441L}.
 The real orbit is slightly eccentric once orbital dissipation starts, hence the instantaneous orbital period 
can have an oscillatory behavior, with the period of oscillations being equal to the real orbital period (see Figure~\ref{fig:Cperiod}). 

We also can find {\it apparent} orbital period  -- this is how long it takes for an observer to see one complete binary revolution.
It is found as follows: 

\begin{enumerate}[(a)]

\item {for the calculations of the apparent orbital period only, 
the center of coordinates $(0,0,0)$ has been assigned to the center of mass of the more massive star; the orbital plane is $X-Y$

\item $t_0$ is the moment of time when the low-mass companion crosses $X-Z$ plane at some $(x_0,0,0)$}

\item $t_1$ is the moment of time when the low-mass companion has passed through 
all the Cartesian quadrants (made $360^o$ rotation around the center of coordinates) and crosses $X-Z$ again at some $(x_1, 0,0)$ 

\item the apparent orbital period is then $P_{\rm orb, app}=t_1-t_0$. 
\end{enumerate}

Note that this method does not intrinsically imply that two interacting 
bodies would necessarily have a Keplerian orbit. 
Due to the effects of tidal bulges, the apparent orbital period is generally smaller than instantaneous period \citep[see for example Equation (7.6) in][]{1993ApJS...88..205L}, especially when the envelope of the primary star is starting to be significantly puffed up 
(see Figure~\ref{fig:Cperiod}). 
Also note that apparent period can be found for the first time only after one orbital period, 
as can be seen in Figure\ref{fig:Cperiod}.
 
\subsection{Ejecta}

\label{sec:def_ej}

For the analysis of our simulations, we define as \textit{ejecta} the unbound 
material of the binary system. We consider two ways to define unbound material.

{\it Conventional definition}. We say that a particle belongs to the ejecta 
if the total energy of that particle (the sum of kinetic, internal and gravitational energies) is positive:

\begin{equation}
\frac{1}{2}m_iv_{i}^2+m_i \Phi_i  +m_iu_i>0.
\label{eq:totalenergy}
\end{equation} 
\noindent Here, $v_i$ is the velocity of the particle $i$ relative to the center of 
mass (fixed at the origin), $m_i$ is the mass of the particle $i$ and $u_i$ 
is the specific internal energy of the particle $i$. 
The potentials of each particle  $\Phi_i$ 
and their gravitational accelerations are calculated using 
direct summation on NVIDIA graphics cards, softened with the usual SPH kernel as in \cite{HK1989}
\citep[for more details on implementation and justification see][]{2011ApJ...737...49L}.
The first term in the equation \eqref{eq:totalenergy} is the kinetic energy, the second 
is the gravitational potential energy, and the third is the internal energy of the particle $i$. 

{\it Abridged definition.}  We say instead that a particle corresponds to the 
ejecta if the sum of kinetic and gravitational energies is positive:

\begin{equation}
\frac{1}{2}m_iv_{i}^2+ m_i \Phi_i >0.
\label{eq:totalenergy2}
\end{equation} 
\noindent  Note that this definition implies that internal energy does not play a role in determining whether the matter leaves the system.
By default in this paper we use the abridged definition, as discussed more below.

\subsection{Common envelope}

\label{sec:rad_ce}

We recognize that the common envelope is formed by the expanding envelope of the primary. 
We define that a particle belongs  to a common envelope if the following conditions are satisfied:

\begin{enumerate}
\item a particle is bound to the binary system -- the sum of kinetic 
and gravitational energies, calculated with respect to the binary, is negative;
\item a particle is located outside of the Roche lobe of the secondary; 
\item a particle is counted only if its density is above the threshold density $\rho_{\rm TR}=10^{-6}$ g~cm$^{-3}$.
\end{enumerate}

At the start of RLOF, material from the primary streams inside the Roche lobe of the secondary, forming an accretion disk. 
Material transferred via the Lagrangian pont $L_1$ can not be unambiguously considered as forming a common envelope, 
at least not before it starts to encompass the Roche lobe of the secondary. 
This motivates the second condition described above, which is taken into account only when 
the orbital separation exceeds the initial primary radius.

At the beginning of the binary interaction, a few low-mass surface particles  
are typically perturbed from the surface but remain bound to the binary for a while. 
However, because they have extremely low densities, 
they do not significantly affect the orbital evolution, nor do they form a continuous envelope. 
Each of those very-low-density particles can take a volume comparable to that of the primary star.
We therefore limited the density of particles that contribute to the common envelope. 
Our threshold density implies that even if a sphere that envelops both the primary and the secondary 
is filled up with particles below this threshold density, the total mass of these particles 
will be $\la 10^{-4}M_\odot$.

For each particle that satisfies the conditions above, we find the volume which that particle fills, $V_i=m_i/\rho_i$. We then sum 
those volumes and solve for the volume-equivalent radius of the common envelope, $R_{\rm CE}$.

\subsection{Entropy}

For each particle, we calculate the entropy as

\begin{equation}
 S_i=\frac{k m_i}{m_{\rm H} \mu_i}\ln \frac{T_i^{3/2}}{\rho_i}+\frac{4}{3}\frac{m_i aT_i^3}{\rho_i}+S_{0,i},
\label{eq:entropypart}
\end{equation}
where $S_{0,i}$ is a constant and depends only on the chemical composition 
of the particle \citep[see][]{2001spfc.book.....B}, 
$T_i$ is the temperature of the particle computed in the same way as in \cite{2006ApJ...640..441L}, 
$\rho_i$ is the density of the particle found by the SPH code, 
$k$ is the Boltzmann's constant, $m_{\rm H}$ is the hydrogen mass, 
$a$ is the radiation constant, and $\mu_{i}$ is the mean molecular weight. 
The SPH code does not evolve the chemical composition -- the code conserves 
$\mu_i$ for each particle in the system. 
However, because the code uses as input realistic stellar models, 
particles can have different $\mu_i$ and accordingly different $S_{0,i}$. 
We find $S_{0,i}$ as described in Appendix~\ref{app:a}.

The specific entropy of the unbound material for each time-step, $s_{\rm unb}$, 
can be obtained by dividing the total entropy of the ejecta 
by the total unbound mass, $m_{\rm unb}$, i.e.
\begin{equation}
 s_{\rm unb} \equiv \frac{\sum_{i, \rm unb} S_i}{\sum_{i, \rm unb} m_i}\ .
 \label{eq:avsentropy}
\end{equation}
\noindent The summation here is only over the unbound particles.

We also define similarly the average temperature of the unbound material at each time-step as

\begin{equation}
\bar{T}_{\rm unb} = \frac{\sum_{i,\rm unb} T_i m_i}{\sum_{i,\rm unb} m_i}\ .
\end{equation}

\subsection{Start of the common envelope, merger, and the end of the simulations}
\label{subsec:merger}

We define several crucial phases in the evolution of our merging binary.

First, we find when the common envelope phase starts, $t_{\rm CE}$. 
For that we use the conventional definition of the common envelope 
-- this is the moment when the companion starts to orbit inside the
 material that is bound to the primary core.
Note that, observationally, a binary likely would not be distinguished 
as a binary from the moment the common envelope phase started. 

When the companion just starts to orbit within the common envelope -- 
the ``loss of corotation''  stage -- the orbit still decays relatively slowly.
This stage is then followed by the plunge-in phase, during which the companion quickly loses its orbital angular momentum 
as the orbit quickly shrinks. 
Assuming that this shrinkage is a half of the orbital separation during one initial orbital period,
the value of $\dot{a}/a$ is about  $-4\times 10^{-6}$ s$^{-1}$. We hence 
adopt the definition $t_{\rm plunge}$ as the time when $\dot{a}/a=-4\times10^{-6}$ s$^{-1}$.

We say that the binary is fully merged when the separation between the cores 
(in other words, the separation in our instantaneous Keplerian orbital solution) 
is less than $0.1\ R_\odot$. This corresponds to an instantaneous orbital period 
\begin{equation}
P_{\rm orb, inst}< 0.004 \left(\frac{M_\odot}{M_{\rm tot}}\right)^{1/2} \rm d, 
\end{equation}
where $M_{\rm tot}$ is the total mass of the binary. (Note that while the companion is already inside the primary star, the
instantaneous orbital period is not a physically valid quantity but is an upper limit for the ``true'', or apparent, period.) 
For our case of $M=1.68\ M_\odot$,  this corresponds to $P_{\rm orb, inst}\lesssim 0.003$ d. 
Thus, the merger time,  $t_{\rm merg}$, is defined as the moment when the orbital 
separation (or the orbital period) is less than $0.1\ R_\odot$ (or $0.003$ d).

\section{Orbital evolution prior to merger.}
 
In this section, we analyze how the initially detached binary approaches RLOF, 
how it starts the common envelope phase, and how it merges.
The ``approach'' phase is far from being well understood 
\citep{2013A&ARv..21...59I}, 
not least because 3D-simulations would usually start at RLOF. 
While many of our models also start close to their RLOF, 
we also simulate a number of cases for dozens of days or more prior the RLOF.
 
\begin{figure}[t]
 \begin{center}
   \includegraphics[scale=0.45]{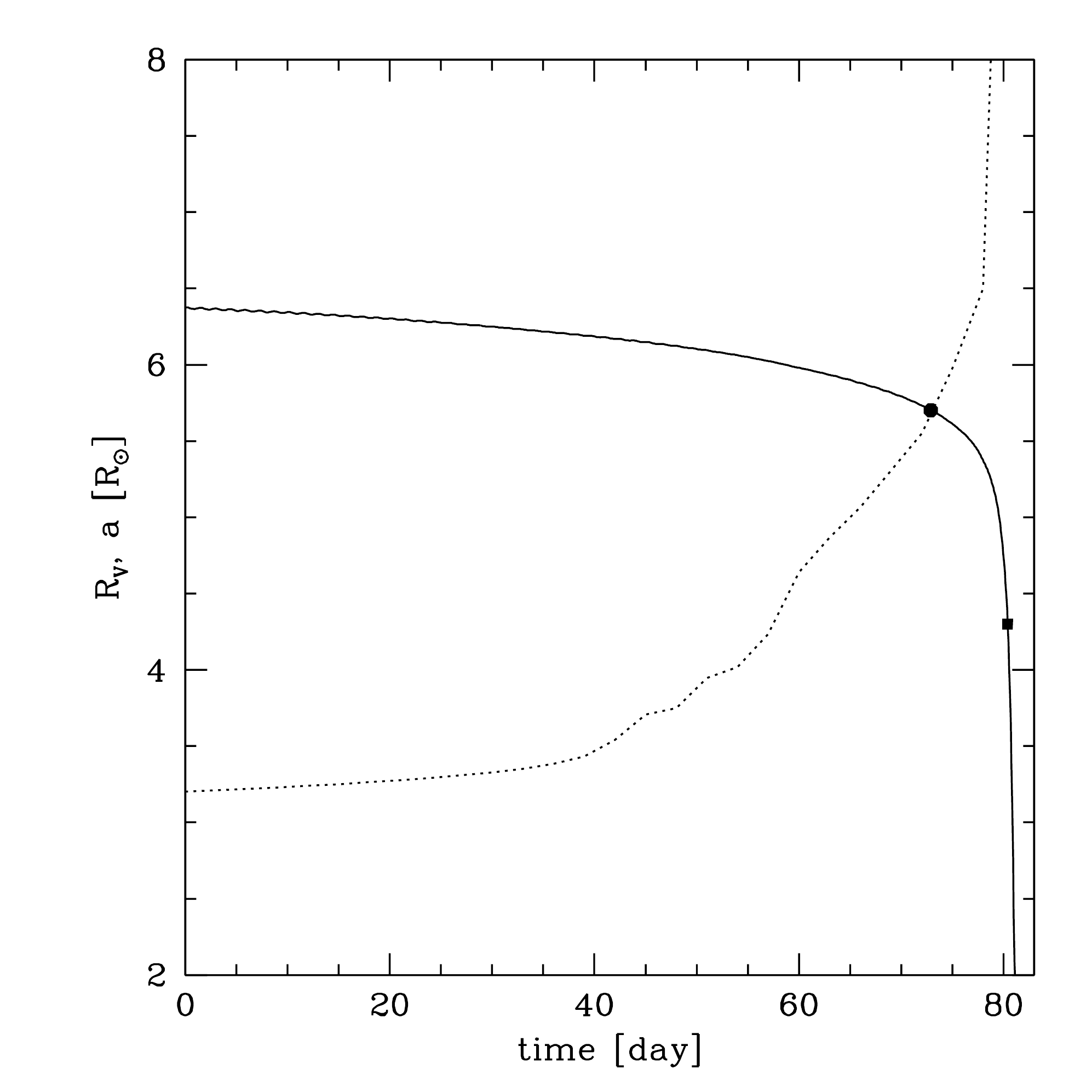}
 \end{center}
\caption{The instantaneous orbital separation (solid line) and 
the radius of the primary star envelope which is transitioning into the common envelope (dotted line) for the simulation pn319. 
The black solid circle marks the start of the common envelope,  $t_{\rm CE}$, 
and black solid box marks the stars of the plunge-in, $t_{\rm plunge}$.}
\label{fig:approach}
\end{figure}

In Figure~\ref{fig:approach}, we show as an example the evolution of the instantaneous 
orbital separation in the model pn319. This separation is compared to the the radius 
of the envelope which initially is just the primary envelope and then transforms to the 
common envelope (the radius of that envelope is calculated as described in the \S\ref{sec:rad_ce}). 
We can distinguish three phases: 

\begin{itemize}
\item the approach to RLOF and the start of the common envelope phase; 
\item the loss of corotation \citep[or  \emph{phase I} of the common envelope event as per adopted classification, see e.g. \S2 in ][]{2013A&ARv..21...59I}; 
\item the plunge-in with the termination (or  \emph{phase II} as per classification).
\end{itemize}

\subsection{Why does the orbit decay and the common envelope start?}

\begin{table}[t]

\caption{Angular momenta and Darwin instability.}
\label{tab:tableang}
\begin{center}
\begin{tabular}{c c c c c c c c}
\hline
 Model & $I_1$& $J_{\rm orb}$ & $J_{\rm cr}$  &$|L_{\rm tot,SPH}|$ & $L_{\rm cr}$ & $L_{\rm b}$ & $L_{\rm unb}$\\
\hline
 ps334 & 1.90 & 2.85 & 2.92  &3.75& 3.83 & 2.76 & 0.987 \\
 mn351 & 2.52 & 2.82 & 3.87  &2.85& 4.10 & 2.35 & 0.502\\
 pn351 & 2.52 & 2.85 & 3.87  &2.85& 4.10 & 2.32 & 0.529\\
 ps351 & 2.06 & 2.85 & 3.17  &3.91& 3.90 & 2.62 & 1.300\\
 ms376 & 2.63 & 2.90 & 3.83  &4.18& 4.15 & 3.25 & 0.937\\
 ps376 & 2.63 & 2.90 & 3.83  &4.18& 4.15 & 3.25 & 0.935\\
 ms372 & 2.60 & 2.87 & 3.92  &4.10& 4.14 & 3.18 & 0.915\\
 ps379 & 2.77 & 2.85 & 4.25  &4.07& 4.20 & 2.92 & 1.150\\
 pn344 & 2.39 & 2.86 & 3.62  &2.87& 4.05 & 2.31 & 0.556\\
 ps375 & 2.64 & 2.86 & 4.00  &4.09& 4.15 & 2.81 & 1.290\\
 mn344 & 2.39 & 2.86 & 3.63  &2.86& 4.05 & 2.37 & 0.494\\
 ms375 & 2.64 & 2.86 & 4.00  &4.09& 4.15 & 3.16 & 0.924\\
 pn319 & 1.84 & 2.86 & 2.80  &2.87& 3.79 & 2.08 & 0.783\\
\hline
\end{tabular}
\begin{tablenotes}
       \item $I_1$ is the moment of inertia of the primary star in $10^{55}$ g cm$^2$,
       \item $J_{\rm orb}$ is the orbital angular momentum of a 2 point-mass binary in $10^{51}$ g cm$^2$ s$^{-1}$,
       \item $J_{\rm cr}$ is the critical orbital angular momentum of a 2 point-mass binary in $10^{51}$ g cm$^2$ s$^{-1}$,
       \item $L_{\rm b}$ is the total angular momentum for the bound material at the end of the simulation in $10^{51}$ g cm$^2$ s$^{-1}$,
       \item $L_{\rm unb}$ is the total angular momentum for the unbound material at the end of the simulation in $10^{51}$ g cm$^2$ s$^{-1}$,
       \item $L_{\rm tot,SPH}$ is the total angular momentum for the SPH particles in $10^{51}$ g cm$^2$ s$^{-1}$, 
       \item $L_{\rm cr}$ is critical angular momentum in $10^{51}$ g cm$^2$ s$^{-1}$.
     \end{tablenotes}
\end{center}
\end{table}

It was suggested by \cite{1879RSPS...29..168D} that if the orbital angular momentum 
of the binary is less than three times the spin angular momentum of its companions, 
the binary is dynamically unstable and the stars would fall to each other. 
The revised condition for the instability is that the configuration is unstable 
once the orbital angular momentum is less than the critical value \citep{1980A&A....92..167H}:

\begin{equation}
J_{\rm orb} < J_{\rm cr} = 3 (I_1 + I_2) \ \Omega \ ,
\label{darwin1}
\end{equation}
\noindent where $I_1$ and $I_2$ are the moments of inertia of the binary components and $\Omega$ is the angular velocity of the synchronous rotation and revolution, $\Omega_{\rm orb}\simeq5\times10^{-5}$ Hz for all our simulations. 
For a detached binary the orbital angular momentum, in the two-point mass approximation, is 
\begin{equation}
 J_{\rm orb}=\sqrt{G\frac{M_1^2M_2^2}{M_1+M_2}a(1-e^2)},
\end{equation}
where  $e$ is the eccentricity of the orbit. All our simulations start in an circular orbit, $e=0$, and
the orbital angular momentum is about the same, $J_{\rm orb} \simeq 2.8-2.9\times 10^{51} {\rm g cm^2/s}$.

We compute the moment of inertia for each of our stars individually. As all rotation is around the $z-$axis, 
\begin{equation}
 I=\sum_i m_i(x_i^2+y_i^2),
\end{equation}
where the particle coordinates $x_i$ and $y_i$ are measured with respect to the center of mass of the star under consideration.
The moments of inertia in our primary stars are in the range $I_1=1.84-2.77\times 10^{55}$ g cm$^2$, and our non-degenerate companion has 
$I_2=1.1\times10^{52}$ g\ cm$^2$. For the subset of 
simulations where the low-mass companion is non-degenerate, the range of $I_1$ is narrower,  $I_1=2.4-2.6\times 10^{55}$~g~cm$^2$.
The smallest value corresponds to the simulation pn319, and the largest value corresponds to the simulation ps379.
In the case of our most compact donor, in the simulation pn319, we have $J_{\rm cr} = 2.8 \times 10^{51}{\rm g cm^2/s}$ (in this simulation, $J_{\rm orb}\simeq 2.8\times 10^{51} {\rm g cm^2/s}$), hence the system is right at the border of the Darwin instability by the criterion defined by the equation~(\ref{darwin1}). 
In all other simulations, $J_{\rm cr} > J_{\rm orb}$.

The other way to express the criterion for the Darwin instability is in terms of the angular momentum: the binary is unstable once the total angular momentum is less than critical value \citep{1980A&A....92..167H}

\begin{equation}
L_{\rm cr} = 4 \left[ \frac{1}{27} G^2 \frac{M_1^3 M_2^3}{M_1 + M_2} (I_1 + I_2)\right]^{1/4} \ ,
\end{equation}
\noindent where $M_1$ and $M_2$ are the masses of two companions (here note that a factor of $G$ was missed in the original work). For our case,

\begin{equation}
L_{\rm cr} \simeq 4.07 \times 10^{51}  \left [ \frac{I_1 + I_2}{2.5\times10^{55}{\rm g\ cm^2}}\right]^{1/4} {\rm g\ cm^2/s} \ ,
\end{equation}

The centre of mass in our simulations is located at the origin. We then compute the total angular momentum of our system by using $\mathbf{L}=\mathbf{r}\times\mathbf{p}$ for each SPH particle:
\begin{equation}
 \mathbf{L}_{\rm tot,SPH}=\sum_i  \mathbf{r}_i \times (m_i\mathbf{v}_i).
\label{eq:ltotsph}
\end{equation}

The condition that $L_{\rm tot}<L_{\rm cr}$ was derived for the case when a binary system is in tidal equilibrium, and tidal equilibrium can be established only if coplanarity, circularity, and corotation have been established \citep[e.g., ][]{1980A&A....92..167H}.
In a binary for which corotation has not yet been established, as in some of our simulations, the instablity sets in even earlier as even more of the orbital angular momentum would have to be spent on spinning up the companions. If a donor star has overfilled its Roche lobe, the condition is also inapplicable, as the system has already become dynamically unstable.
Table~\ref{tab:tableang} shows that $L_{\rm tot,SPH}<L_{\rm cr}$ in all our simulations, except when $R_*$ is significantly larger than the Roche lobe radius.  This re-confirms that the system we consider is affected by Darwin instability. 

\begin{figure}[t]
 \begin{center}
   \includegraphics[scale=0.45]{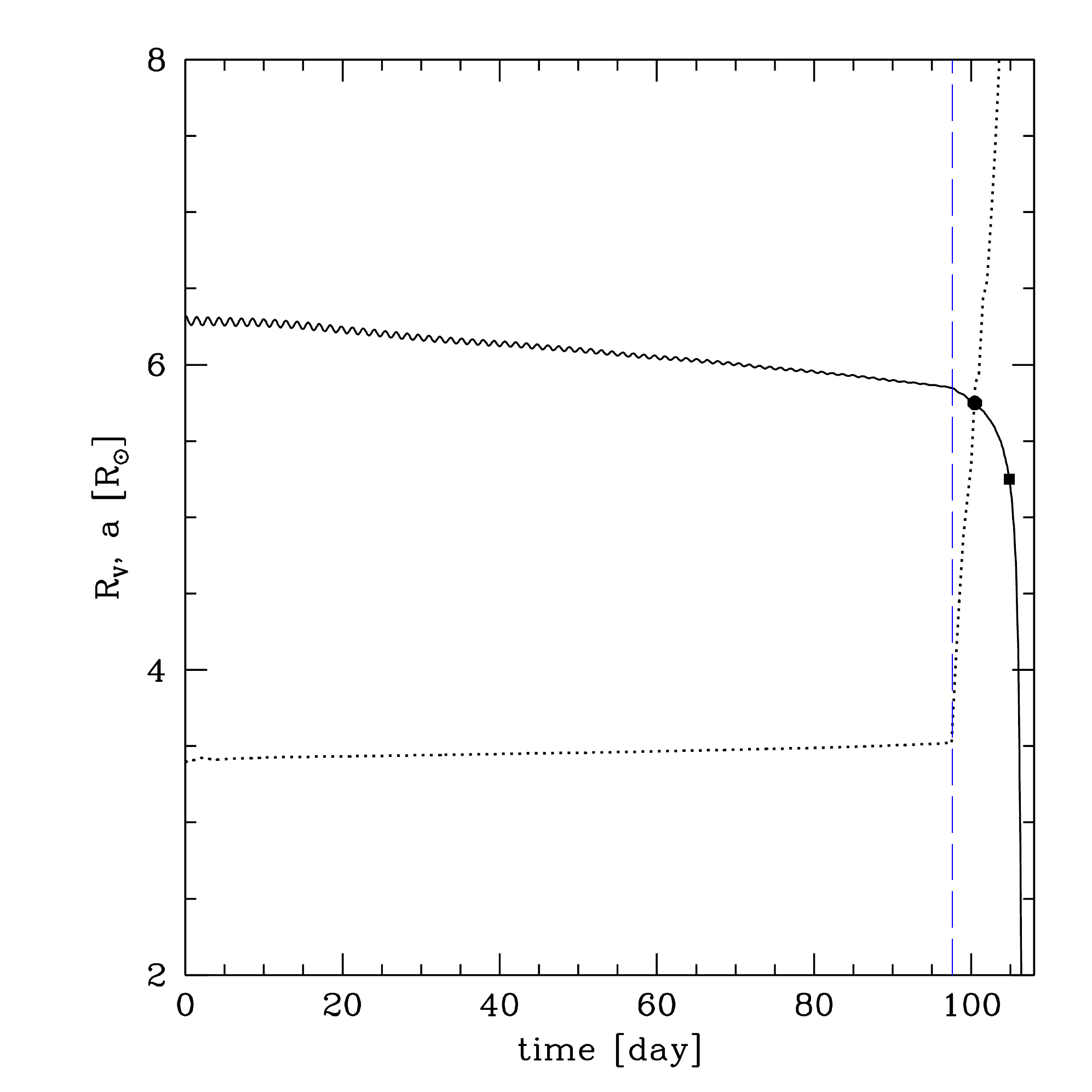}
 \end{center}
\caption{The instantaneous orbital separation (solid line) and 
the radius of the primary star envelope transitioning into the common envelope (dotted line) for the simulation ps351.
The blue dashed line marks 97.6 days, when the sharp decay in the orbit and rapid increase in the radius of the primary start. 
The black solid circle marks the start of the common envelope,  $t_{\rm CE}$, 
and black solid box marks the stars of the plunge-in, $t_{\rm plunge}$.}
\label{fig:approachps351}
\end{figure}

\subsection{Synchronization of the binary system and how the primary expands.}

\begin{figure*}[t]
 \begin{center}
   \includegraphics[scale=0.9,angle=-90]{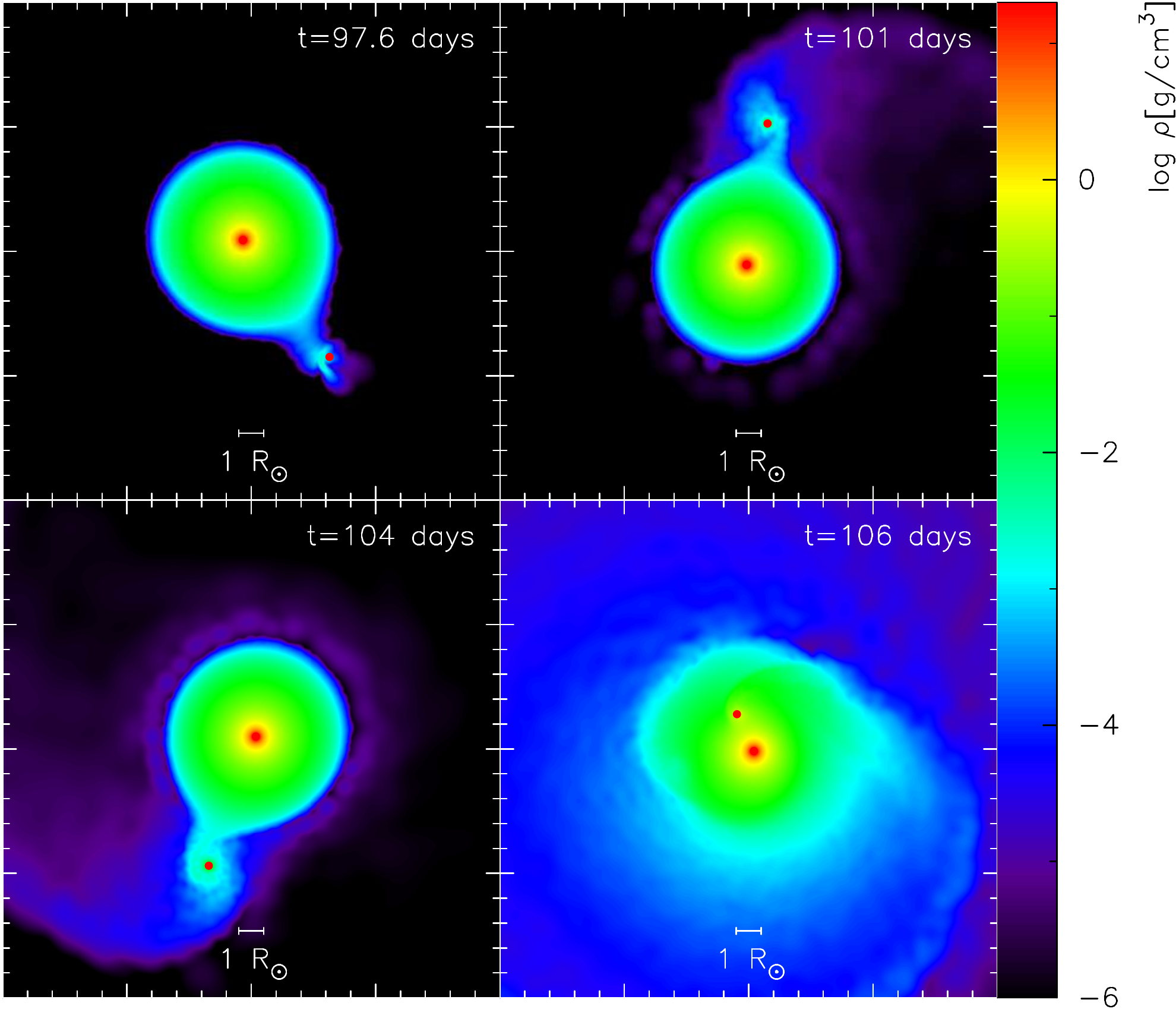}
 \end{center}
\caption{Cross-sectional slices for density in the orbital plane for the simulation ps351.  The top left panel ($t=97.6$ d) is when the primary has overflown its Roche lobe and more material is being transfered to the companion. The top right panel ($t=101$ d) shows that the Roche lobe of the companion is overflown. The bottom left panel ($t=104$ d) is for the stage when the companion spirals into the primary, while the bottom right panel ($t=106$ d) shows the two orbiting cores engulfed by the envelope of the primary; after about $0.9$ d the cores merge. }
\label{fig:ps351RLOF}
\end{figure*}

Let us consider first the model ps351, with a synchronized donor and a degenerate companion.  The simulation
starts with a primary that has $f_{\rm RLOF}<1$ but $f^*_{\rm RLOF}>1$.
Once the primary has filled the volume equivalent of its Roche lobe, 
its surface material starts to expand rapidly into the Roche lobe of the companion
(see Figure~\ref{fig:approachps351} and the top left panel 
in Figure~\ref{fig:ps351RLOF}).
After this moment, the primary keeps expanding, overfilling its Roche lobe. 
Only 2.8 days elapse between the initial RLOF and the common envelope formation,
even though the CE does not appear very well visually 
distinguished in the  right top panel of Figure~\ref{fig:ps351RLOF}.
The time between the primary starting to expand rapidly 
and the moment when the common envelope has formed
is only two initial orbital periods -- this is a dynamical event.

Another synchronized model, the model ps334, 
is a binary where the donor was well inside of its Roche Lobe as even $f^*_{\rm RLOF}<1$. However, this model 
also shows the same characteristic behavior described for the model ps351  -- in particular, it shows 
the same rapid expansion of the primary once RLOF commences. 
We find that this fast increase of the primary radius with the CE starting soon thereafter is observed in all 
synchronized simulations with a degenerate companion.
The same models show that most of the transferred mass is lost from the binary via the Lagrangian point $L_2$.
The duration of mass loss through $L_2$ is comparatively short -- e.g., in ps351 $L_2$ mass loss starts at about 97.7 days, 
just after the rapid radius increase starts: $L_2$ mass loss occurs only 
 when a dynamical timescale mass transfer takes place.

Note that this rapid expansion does not necessarily lead to the drastic changes in the light-curve of the outburst, which could be solely formed by the recombination wave fronts, \cite{2013Sci...339..433I}.

\begin{figure*}[t]
 \begin{center}
   \includegraphics[scale=0.9,angle=-90]{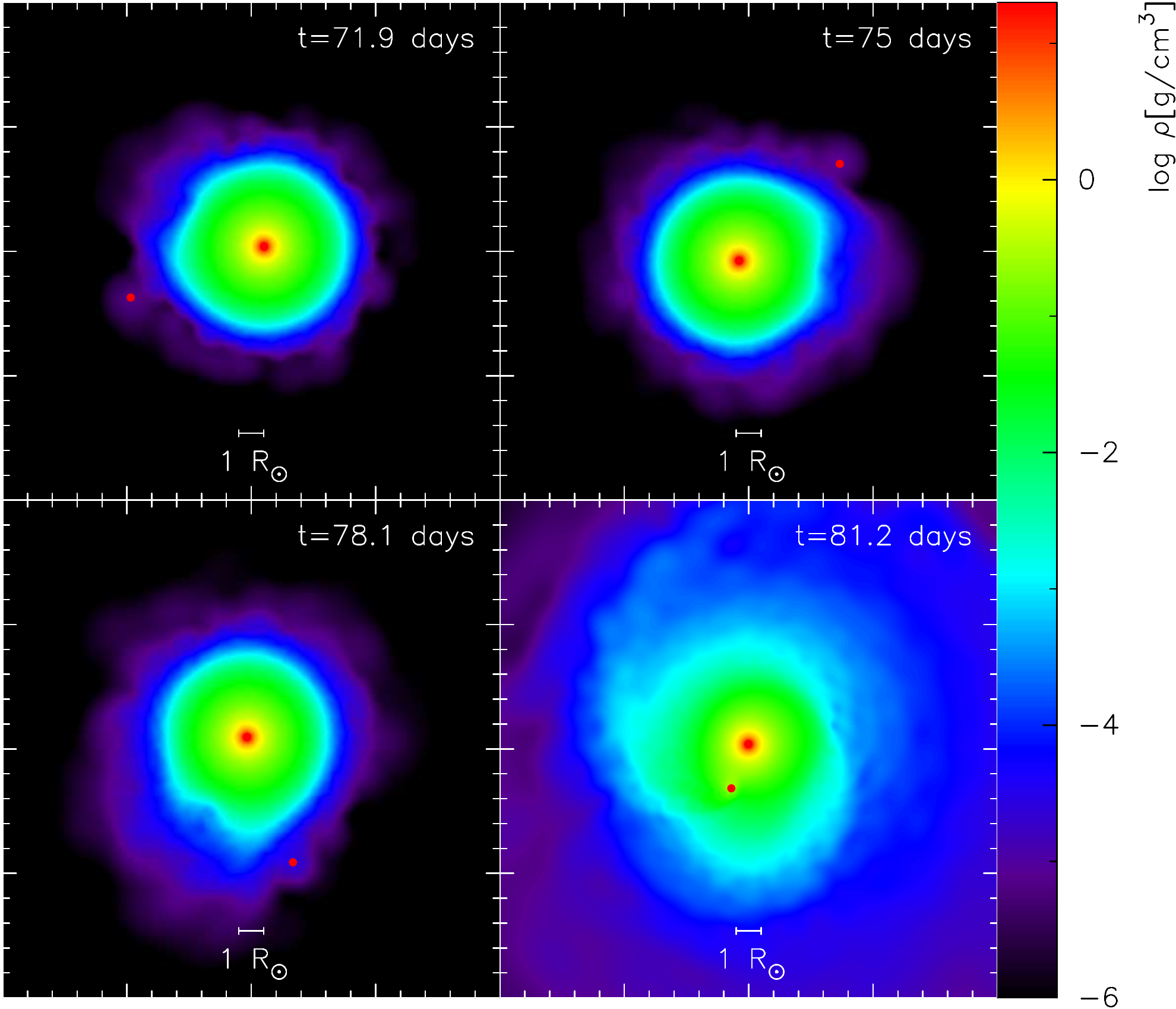}
 \end{center}
\caption{Cross-sectional slices for density in the orbital plane for the simulation pn319.  The top left panel ($t=71.9$ d) is when the primary has overflown its Roche lobe and more material is being passed to the companion. The top right panel ($t=75$ d) shows that the Roche lobe of the companion is overflown. The bottom left panel ($t=78.1$ d) is for the stage when the companion spirals into the primary, while the bottom right panel ($t=81.2$ d) shows the two orbiting cores engulfed by the envelope of the primary; after about $0.5$ d the cores merge. }
\label{fig:pn319RLOF}
\end{figure*}

Now let us consider the case where the donor is not synchronized at the start of a simulation, 
e.g. pn319. Like ps334, this system started with the primary well within its Roche lobe and
the donor starts to transfer mass when it overflows its Roche lobe, on day 57.2 in Figure~\ref{fig:approach}, and 
see also Figure~\ref{fig:pn319RLOF}. 
The star keeps slowly expanding while overfilling its Roche lobe, until it reaches the size of the orbit, then
the fast mass transfer starts to the interior of the Roche lobe of the secondary. 
A slow increase in radius is observed in all non-synchronized models as well as in synchronized with a non-degenerate companion.
In none of the non-synchronized models do we observe any noticeable $L_2$ mass loss -- the material from the Roche lobe of the secondary, if lost, 
was lost isotropically.

We speculate that this slow expansion of the primary star towards the orbit could be a numerical artifact related to SPH particles oscillations around their positions by a smoothing length $h$ (about 5\% of the radius for surface particles). For the most of the simulated models it implies that the primary star would find itself often in ``instantaneous'' RLOF; this unavoidably speeds up the start of the common envelope phase. On the other hand, the oscillations by 5\% of the stellar radius are comparable to the scale over which convective eddy exists. In giants that have surface gravity close to zero, the surface of a giant is not smooth, the convective plums, which are comparable in size to the giant radius, would rise above the conventionally defined surface \citep[e.g., see ][]{2011A&A...528A.120C}.

\begin{table}[t]

\caption{Important times.}
\label{tab:tablerestime}
\begin{center}
\begin{tabular}{c  l l l l}
\hline
 Model  &$t_{\rm CE}$&$t_{\rm plunge}$ & $t_{\rm merg}$ & $t_{\rm end}$\\
\hline
 ps334  &154.91  &157.10 &  158.4 &283 \\
 mn351  &10.40   &13.42&  13.6  & 15  \\
 pn351  &11.80   &13.08&  14.7  &30   \\
 ps351  &100.45  &104.85&  106.9 &156 \\
 ms376  & 26.9   &29.02 &  31.2  & 43 \\
 ps376  & 39.72  &42.58 &  44.9  & 61 \\
 ms372  & 11.2   &12.80&  15.1  & 23 \\
 ps379  &3.30    &6.28&  7.7   &55  \\
 pn344  &21.40   &24.01&  25.6  &46  \\
 ps375  &8.90    &12.74&  14.1  &36  \\
 mn344  &20.50   &24.22&  24.4  &31  \\
 ms375  &9.66    &13.50&  13.8  &20  \\
 pn319  &72.90   &80.40&  81.7  &115 \\
\hline
\end{tabular}
\begin{tablenotes}
        $t_{\rm CE}$ is when the low-mass companion is engulfed by the envelope of the more massive star, $t_{\rm plunge}$ is when the low-mass companion is plunged into the donor, $t_{\rm merg}$ is when the merger took place and $t_{\rm end}$ is last moment in the simulations. The times are in days.
     \end{tablenotes}
\end{center}
\end{table}

\begin{figure}[h]
 \begin{center}
   \includegraphics[scale=0.45]{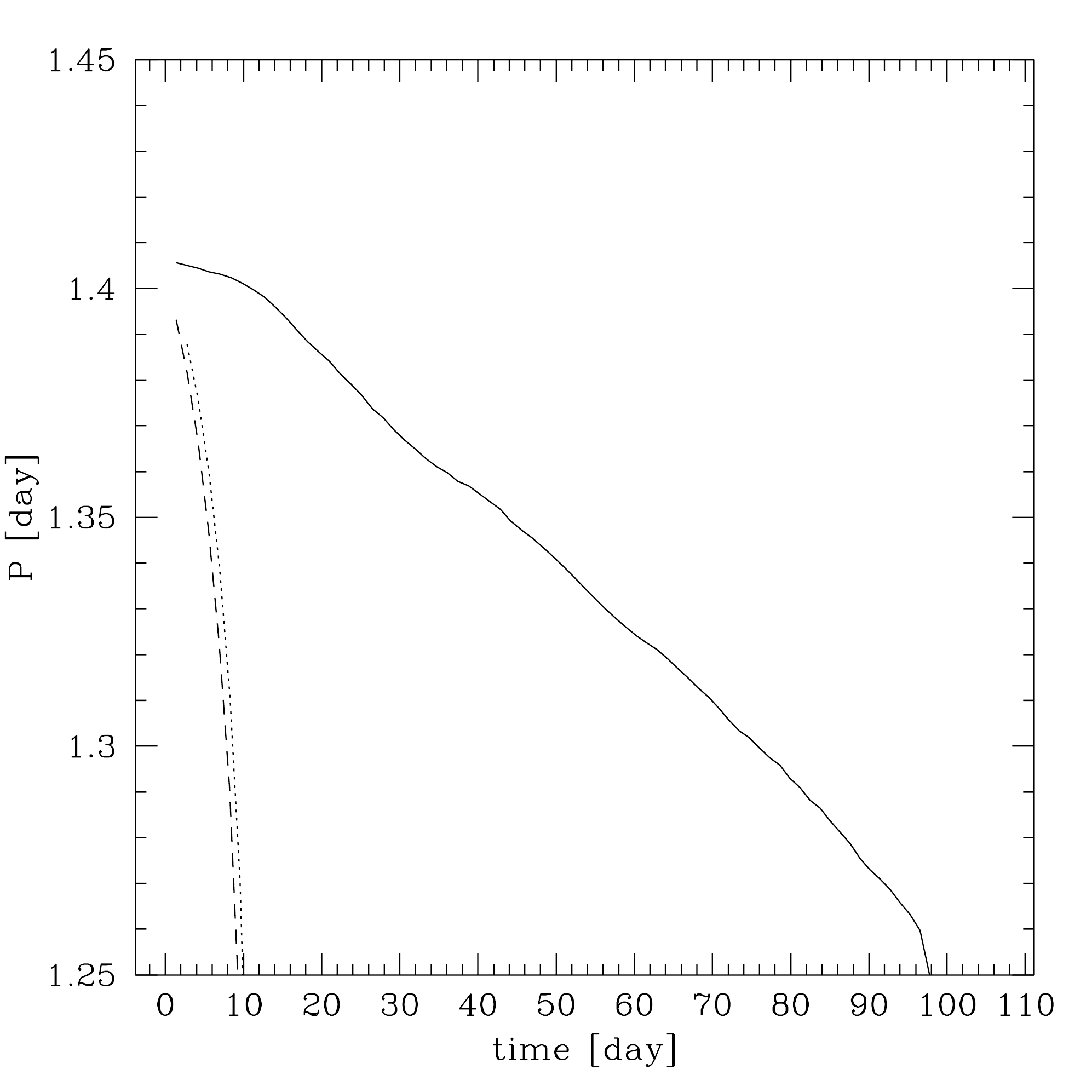}
 \end{center}
\caption{Evolution of the orbital period for the simulations pn351 (dotted), ps351 (solid), and mn351 (dashed).}
\label{fig:Dperiod}
\end{figure}

\subsection{Synchronization of the binary system and the timescale}
To understand how the initial conditions affect the pre-merger evolution, we first consider the effect of synchronization. There are several pairs of simulations that have the same initial conditions except for $f_{\rm sync}$:

\begin{enumerate} [{I.}] 
\item pn351 and ps351 both have primaries at about RLOF (see also Figure \ref{fig:Dperiod}).
\item ps334 and pn319 have almost identical conditions and both primaries are well within their RLOF.
\item two pairs pn344-ps375 and mn344-ms375.  These pairs are harder to analyze cleanly.  In each pair, the relaxed primary has significant RLOF in the case of a synchronized binary, while the non-synchronized binary is well within its Roche lobe.
\end{enumerate}

It can be seen, that if the relaxed stars are within their Roche lobes ($f^*_{\rm RLOF}<1$),  then the synchronized binary, as expected, has a slower period decay, with up to a 10 times difference (see table \ref{tab:tablerestime} data for the simulations in pairs I and II above). Being an RLOF binary cancels this effect, and in case III -- synchronized but overflowing their Roche lobe primaries with $f^*_{\rm RLOF}>1$, -- results in a faster merger.

We conclude that the synchronization prior to Roche lobe overflow leads to a slower period decay.

\label{per_sync}

\subsection{Companion's nature and the timescale}
We consider the effect of the companion's nature, degenerate (represented by a point) vs non-degenerate (represented by a group of particles)
by comparing the following pairs:
\begin{enumerate}[I.]
\item pn344 and mn344  (primaries are well within their Roche lobes, $f^*_{\rm RLOF}<1$)
\item pn351 and mn351 ($f^*_{\rm RLOF}<1$ but $f^*_{\rm RLOF}=1.02$ -- primaries have some particles going beyond their Roche lobes)
\item ps376 and ms376  ($f^*_{\rm RLOF}=1.05$ )
\item ps375 and ms375 (primaries are near their Roche lobe limit with $f_{\rm RLOF}=0.99$ and $f^*_{\rm RLOF}=1.08$)
\end{enumerate}

Unlike the comparison in \S\ref{per_sync}, the degree of RLOF in the primary is the same within each pair, as each member of the pair has the same synchronization. 

In binaries where the primary is well within its Roche lobe (case I) or is just at its Roche lobe limit (case II), the merger time only weakly depends on the nature of the companion, differing only by about a day (see Table~\ref{tab:tablerestime}).
In cases of larger primaries that also noticeably overfill their Roche lobes (cases III and IV), a non-degenerate companion leads to a shorter merger timescale, by up to 3 times. We can conclude that a non-degenerate companion may affect the orbital decay timescale but likely not as significantly as the synchronization of companions.

\subsection{Can we match the observations?}

\begin{table}[t]

\caption{Fit parameters for simulations with merger times larger than $20$ days.}
\label{tab:tableper}
\begin{center}
\begin{tabular}{cccclr}
\hline
 Model &$P_0$&$b_0$&$t_0$&$RSS$&$\nu$\\
\hline
 ps334  &$1.91\pm 0.02$&$135\pm 10$&$440\pm10$&$114$&85 \\
 ps351 &$2.17\pm0.08$&$202\pm30$&$470\pm30$&$169$&69\\
 ms376 &$1.537\pm0.003$&$1.52\pm0.07$&$36.0\pm0.4$&$30$&16\\
 ps376 &$1.84\pm0.04$&$35\pm6$&$159\pm10$&$10$&21\\
 pn344 & $1.506\pm0.002$&$1.49\pm0.03$&$28.3\pm0.1$&$6$&12\\
 mn344 & $1.500\pm0.001$&$1.33\pm0.02$&$26.8\pm0.1$&$3$&11\\
 pn319 & $1.547\pm0.002$&$7.8\pm0.1$&$105.5\pm0.5$&$22$&46\\
\hline
\end{tabular}
\begin{tablenotes}
       \item $P_0$, $b_0$ and $t_0$ are the fitted parameters for the function given by the eq.~(\ref{eq:decay}) (in days);
       \item $RSS$ is the residual sum of squares in units of $10^{-6}$.
       \item $\nu=N-n$ is the number of degrees of freedom, where $N$ is number of observations and $n$ is the number of fitted parameters. Only models with $\nu > 10$ are shown.
     \end{tablenotes}
\end{center}
\end{table}

\cite{2011A&A...528A.114T} found that they can fit V1309~Sco pre-outburst observations
with an exponential period decay, as a function of time $t$:

\begin{equation}
 P_{\rm obs}=P_0 \exp \left(\frac{b_0}{t-t_0}\right)\ ,
\label{eq:decay}
\end{equation}
\noindent  where $P_0=1.4456$ (the period in days), $b_0=15.29$ and $t_0=2455233.5$ was a Julian Date at several hundred days after the merger took place.  The binary period decay was traced in observations for about 2000 days before the binary was seen last as such. 
Numerical simulations of a binary that is almost at its Roche lobe overflow and for the duration of thousands of its orbital periods is well beyond both numerical capabilities of our code and the computational time demand. In our longest simulation, the binary spent 158 days before it merged. A similarly short interval prior to merger has only four observational data points for periods, where each of those four points was derived using 50 observations; the errors in the period determination for those four points were $0.002-0.008$d.  We therefore can attempt to qualitatively compare only the tail of the decay, while assuming that the same fit is valid for the tail of observations as for the whole set.

For each simulation that started with $f_{\rm RLOF} < 1$, we trace the apparent orbital period decay from the start of the simulation until the start of the common envelope phase, $t_{\rm CE}$. 
We fit this orbital evolution to the exponential decay described by Equation~(\ref{eq:decay}) in order to find best-fit values for $P_0$ and  $b_0$; we also look at how quickly the system completes the merger, $t_{\rm merg}$.  The results for all the simulations that have been evolving for 20 days or more before the binary has merged are shown in Table \ref{tab:tableper}.

\begin{figure}[t]
 \begin{center}
   \includegraphics[scale=0.45]{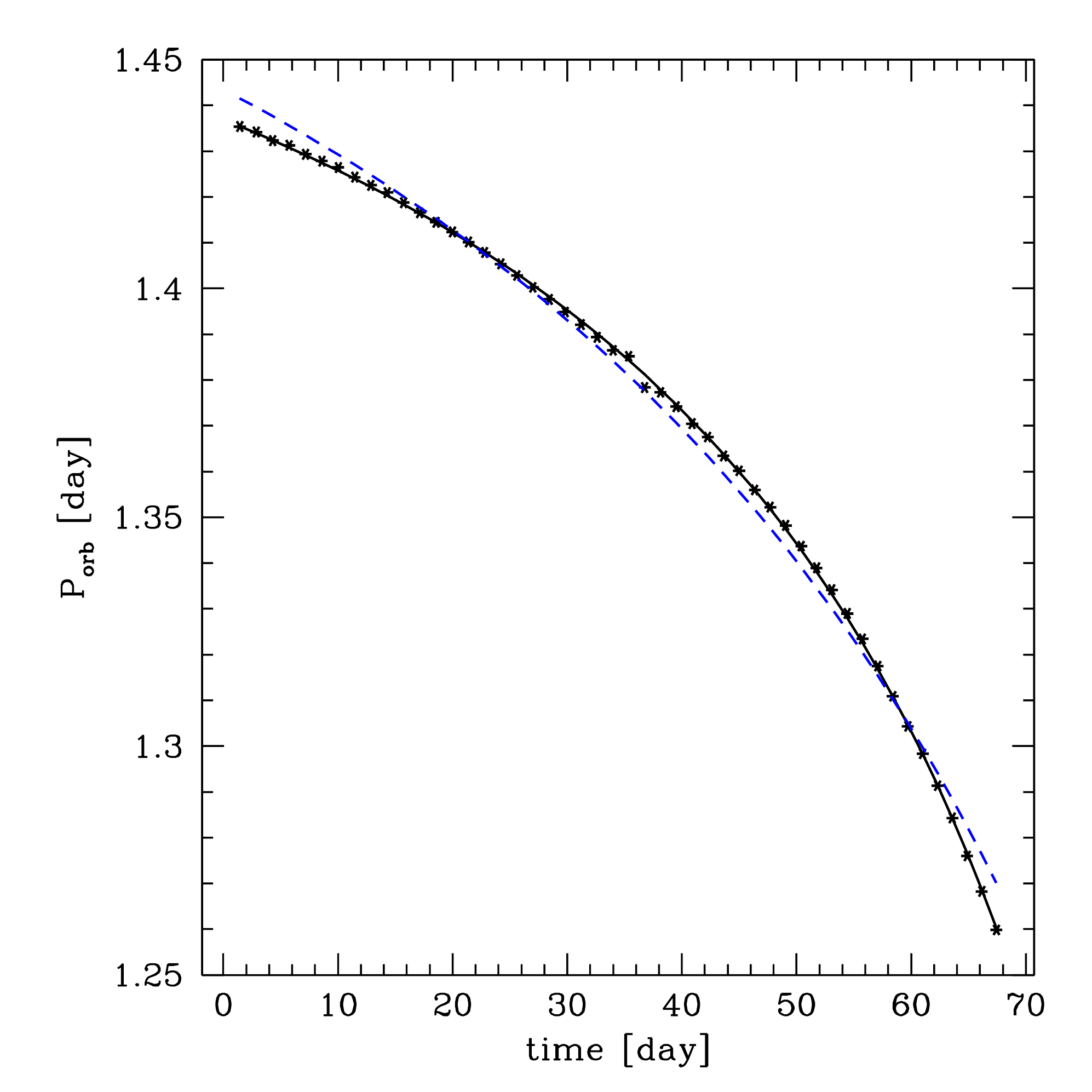}
 \end{center}
\caption{Evolution of the orbital period for the simulations pn319 (the dots). The solid line is the best fit with  $P_0=1.5472$ d, $t_0=105.458$ d, and $b_0=7.7985$ d, while the blue dashed line is the fit with $P_0=1.6242$ d, $t_0=129.596$ d, and $b_0=15.29$ d.}
\label{fig:periodpn319}
\end{figure}

Equation~\ref{eq:decay} implies that an exponential orbital decay takes place if $b_0$ is much smaller than $t_0$. 
We find that the models ms376, pn344, mn344 and pn319 have the period decay shaped similarly to that found by \cite{2011A&A...528A.114T} -- 
Table   \ref{tab:tableper} shows that on those simulations $t_0$ is much larger that $b_0$. 
On the other hand, for a monotonical period decay, $b_0$ should be of the order of $t_0$. 
In the models ps334, ps351 and ps376, $t_0$ is about 3 times larger than $b_0$, and 
the decay in those simulations is almost linear with time, unlike in the observations of V1309~Sco. 

In Figure~\ref{fig:Dperiod} we show examples of the linear and exponential orbital decays in the models ps351, pn351 and mn351 
(note that pn351 and mn351 did not have enough pre-merger models to deduce values of $b_0$ and $t_0$).
Note that in the simulation ps351, the orbital period 
decay is a linear decay for about 90 days, with an abrupt decline thereafter.

In Figure~\ref{fig:periodpn319} we show two fits for Equation ~\ref{eq:decay} for the simulation pn319. 
One fit uses values from \cite{2011A&A...528A.114T}, and another uses fit parameters as in Table \ref{tab:tablesol}. 
The difference between the results is marginal and within the error bar from observations ($\la 0.01$d).
We conclude that the period decay of this model can be fit with about the same shape as the period decay found for V1309 Sco.

We conclude that even though synchronized systems with a degenerate companion have a longer period decay time, they do not exhibit the {\it shape} of the decay observed in the case of V1309 Sco.  This shape can be explained by either (i) a nonsynchronized binary with a degenerate companion or (ii) a binary with a main sequence companion (either synchronized or unsynchronized).

\section{Ejecta}

\subsection{Which material is unbound?}

\begin{figure}[t]
 \begin{center}
   \includegraphics[scale=0.45]{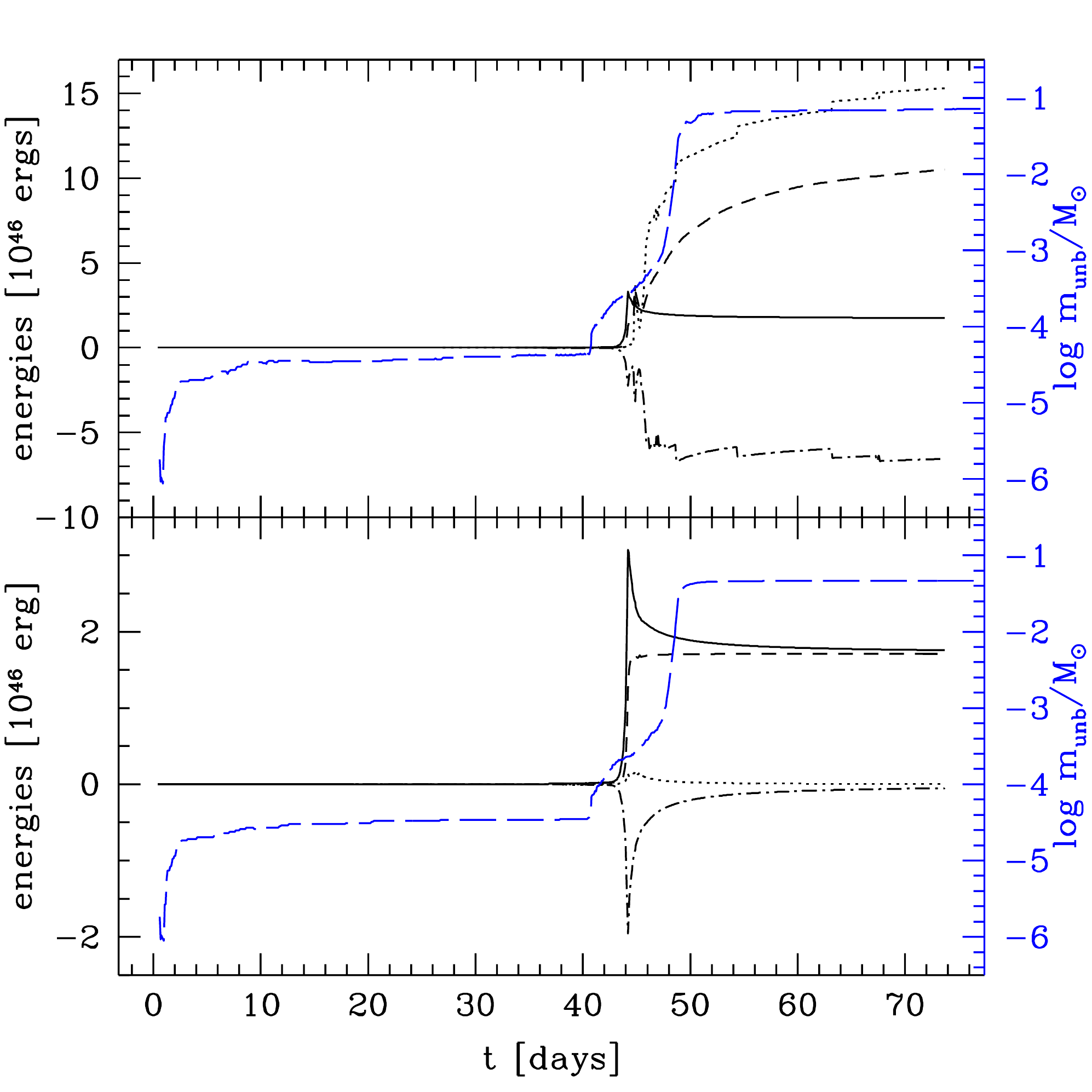}
 \end{center}
\caption{The energies in the ejecta in the simulation ps376 -- 
the kinetic energy (solid line), the internal energy (dotted line), 
the potential energy (dash-dotted line), and  the total energy (dashed line); whilst the blue long-dashed line shows the evolution of the ejecta mass.
On the top panel the ejected material is determined using the criterion \eqref{eq:totalenergy}, 
and on the bottom panel the ejected material is determined using the  criterion \eqref{eq:totalenergy2}. }
\label{fig:internalenergy}
\end{figure}

It has been proposed in \cite{2013Sci...339..433I} that the outburst of V1309~Sco was controlled by the recombination of material ejected during the binary merger. The total energy of the outburst, during the recombination, would therefore depend on the amount and speed of the ejected material. To reproduce the light-curve of the outburst one needs then the ejecta mass loss rate as a function of time. In this section we concentrate on the details of how to recover the mass loss rate with time. This
task requires the identification of the ejected material {\it right at the moment when it starts its initial escape}. 

 In \S\ref{sec:def_ej} we have discussed two ways to define the unbound material. Let us consider how both definitions work in the case of some particular example, the simulation ps376.
In Figure~\ref{fig:internalenergy} we show the evolution of the kinetic, potential, internal, and total 
energies for the material that was classified as ejecta using the criteria given by Equations~\eqref{eq:totalenergy} 
and \eqref{eq:totalenergy2}.

In the ``conventional'' case, the internal energy of all ``ejected'' material 
greatly exceeds its kinetic energy (see Figure~\ref{fig:internalenergy}). 
As the simulation proceeds, the internal energy stays at a large value. 
This is not what would be expected  in a case of an adiabatic expansion anticipated for our ejected and expanding material. 

With the ``abridged'' definition, the internal energy decreases with time as expected for an adiabatic expansion. 
In this case the kinetic energy dominates the energy of the ejecta by the end of the simulation,
even though values of the kinetic energy in the ejected material by both definitions are similar. 

A careful check shows that the difference between the two methods is primarily due to several 
particles located around the low-mass companion. 
The internal energy of these shock-heated particles is high, but their relative velocity to the center of mass is very low.

\begin{figure}[t]
 \begin{center}
   \includegraphics[scale=0.45]{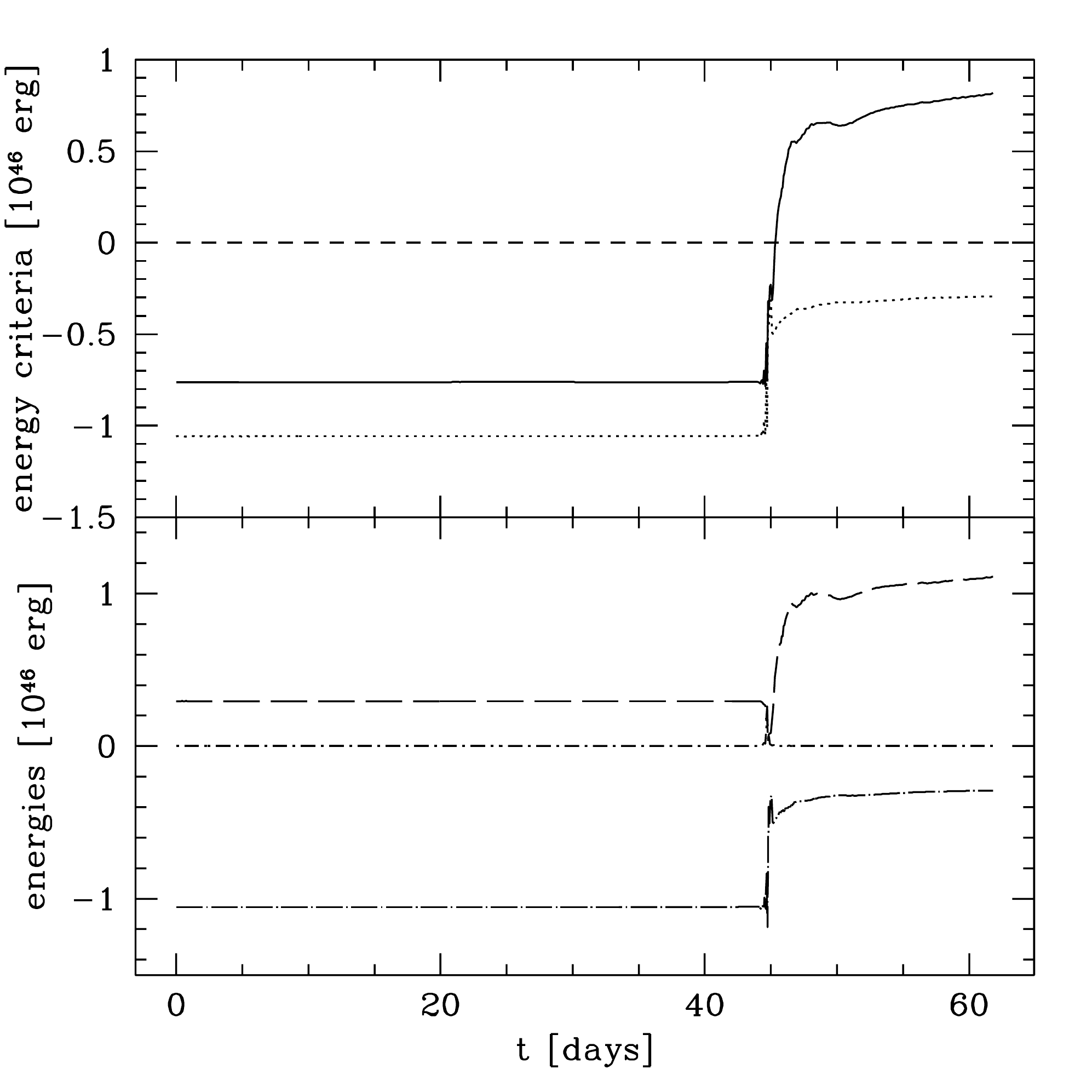}
 \end{center}
\caption{The top panel: how two criteria on determining unboundedness work for the particle 52287 in the simulation ps376 (solid line -- the conventional definition, dotted line -- the abridged criterion). The bottom panel: the kinetic (dotted-dashed line), potential (long dotted-dashed line), and internal (long dashed line) energies for the same particle. }
\label{fig:part52287}
\end{figure}

Figure~\ref{fig:part52287} compares how both criteria work for the particle 52287 in the simulation ps376. 
It can be clearly seen that the internal energy of this particle at all time is much larger than its kinetic energy. 
After the binary merges, the ``conventional'' criterion implies that this particle is unbound to the system, 
while the ``abridged'' criterion indicates that the particles is bound.

\begin{figure}[t]
 \begin{center}
   \includegraphics[scale=0.45]{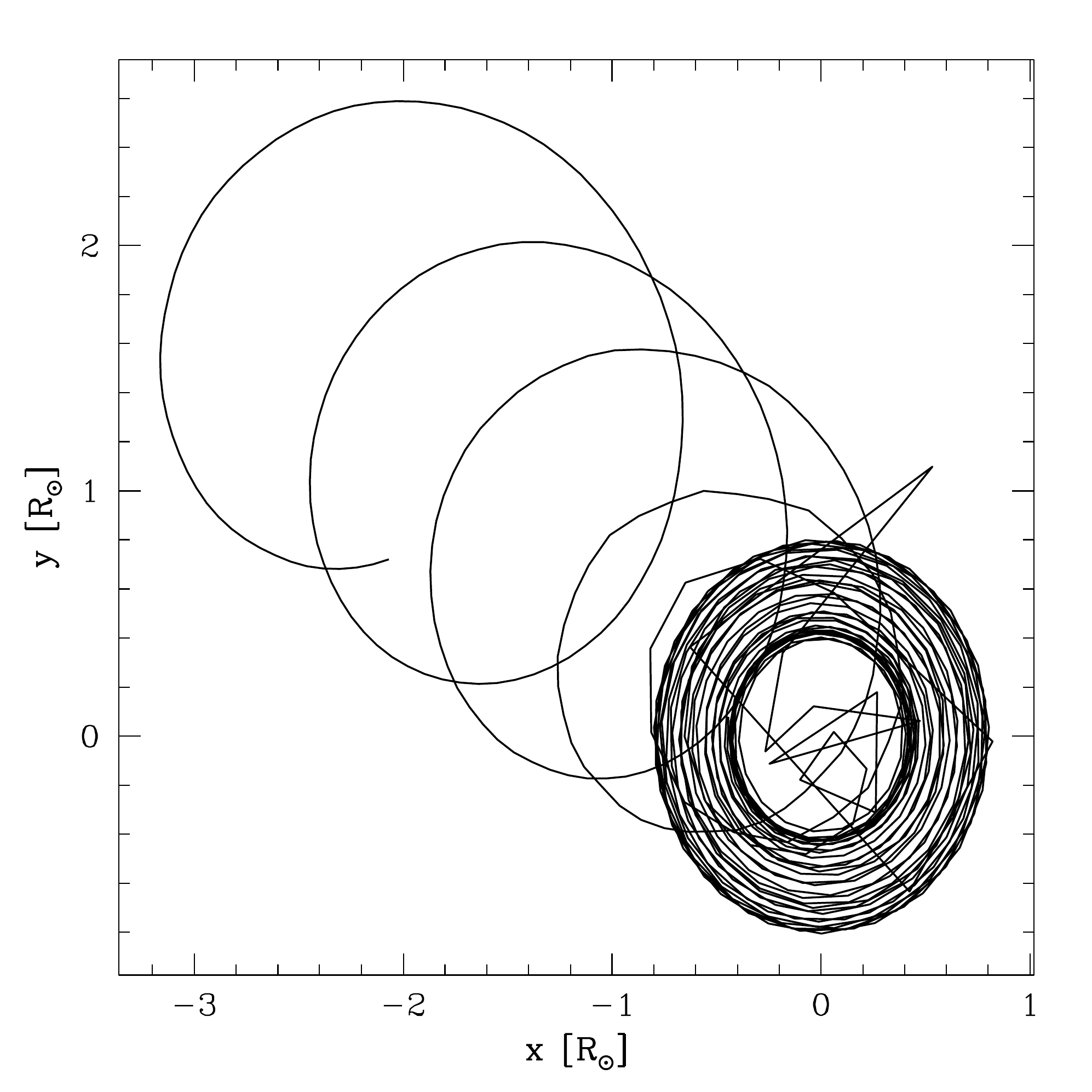}
 \end{center}
\caption{Trajectory of the particle 52287 in the simulation ps376 projected onto the orbital plane.  The particle finishes near $(x,y)=(-2.1,0.7)$ at the end of the simulation.}
\label{fig:orb52287}
\end{figure}

In Figure~\ref{fig:orb52287} we show the positions of the particle 52287 projected onto the orbital plane. 
It can be seen that this particle remains in the vicinity of the merger product, 
its orbit becomes wider after the merger, but its path follows the movement of the center of mass of the merger product until the end of the simulation.
We find that this particle (among other similarly heated particles) is unable to transfer its 
heat to neighboring SPH particles on the timescale of the simulation. 
Therefore, we chose to use the ``abridged'' criterion in order to classify all the particles with this behavior as bound particles. 

\subsection{Mass outbursts.}

\begin{table*}[t]
\caption{Mass, duration and kinetic energy of each episode of mass for each simulation.}
\label{tab:permass}
 \centering
 \begin{tabular}{c|ccc|ccc|ccc|ccc}
  \hline
  Model&$m_{\rm unb}^{b}$&$m_{\rm unb}^{d}$&$m_{\rm unb}^{a}$&$t^{b}$ &$t^{d}$&$t^{a}$&$E_{\rm kin,\infty}^{b}$ &$E_{\rm kin,\infty}^{d}$&$E_{\rm kin,\infty}^{a}$&$E_{\rm kin,max}^{b}$ &$E_{\rm kin,max}^{d}$&$E_{\rm kin,max}^{a}$\\
  \hline
  ps334&--&0.0405&0.0048&--&4.02&1.15&--&1.62&0.08&--&5.91&0.15\\
  mn351&0.0100&0.0280&--&0.97&2.63&--&0.53&0.95&--&1.40&1.34&-- \\
  pn351&0.0115&0.0160&0.0123&1.08&1.44&5.92&0.59&0.55&0.14&1.69&1.04&0.23\\
  ps351&--&0.0467&0.0227&--&3.80&5.50&--&1.85&0.24&--&6.19&0.45\\
  ms376&0.0321&0.0123&--&1.90&1.6&--&1.08&0.25&--&4.06&0.40&--\\
  ps376&--    &0.0409&--&--&2.28&--&--&1.54&--&--&5.40&--\\
  ms372&0.0317&0.0119&--&2.10&1.50&--&1.17&0.32&--&4.27&0.43&--\\
  ps379&--&0.0410&0.0255&--&3.70&3.10&--&1.65&0.38&--&6.27&0.65\\
  pn344&0.0128&0.0165&0.0085&1.30&1.25&4.38&0.64&0.56&0.09&2.13&1.08&0.16\\
  ps375&--&0.0412&0.0384&--&2.70&4.30&--&1.61&0.56&--&6.12&0.90 \\
  mn344&0.0108&0.0237&--&1.00&2.00&--&0.52&0.62&--&1.52&1.09&-- \\
  ms375&0.0320&0.0130&--&2.00&1.40&--&1.22&0.33&--&4.45&0.44&-- \\
  pn319&0.0147&0.0199&0.0209&1.20&2.40&8.40&0.67&0.73&0.31&2.19&1.33&0.51\\
\hline
 \end{tabular}
\begin{tablenotes}
 \item $m_{\rm unb}$, $t$, $E_{\rm kin,\infty}$ and $E_{\rm kin,max}$ are the mass ejecta in $M_\odot$, duration in days and kinetic energy at infinity in $10^{46}$ ergs and maximum kinetic energy at the moment of ejection of each mass ejection. The superscript $a$ implies ``after the merger'', and the subscript $b$ implies ``before the merger'' while $d$  implies ``during the merger''.
\end{tablenotes}

\end{table*}

\begin{figure}[t]
 \begin{center}
   \includegraphics[scale=0.45]{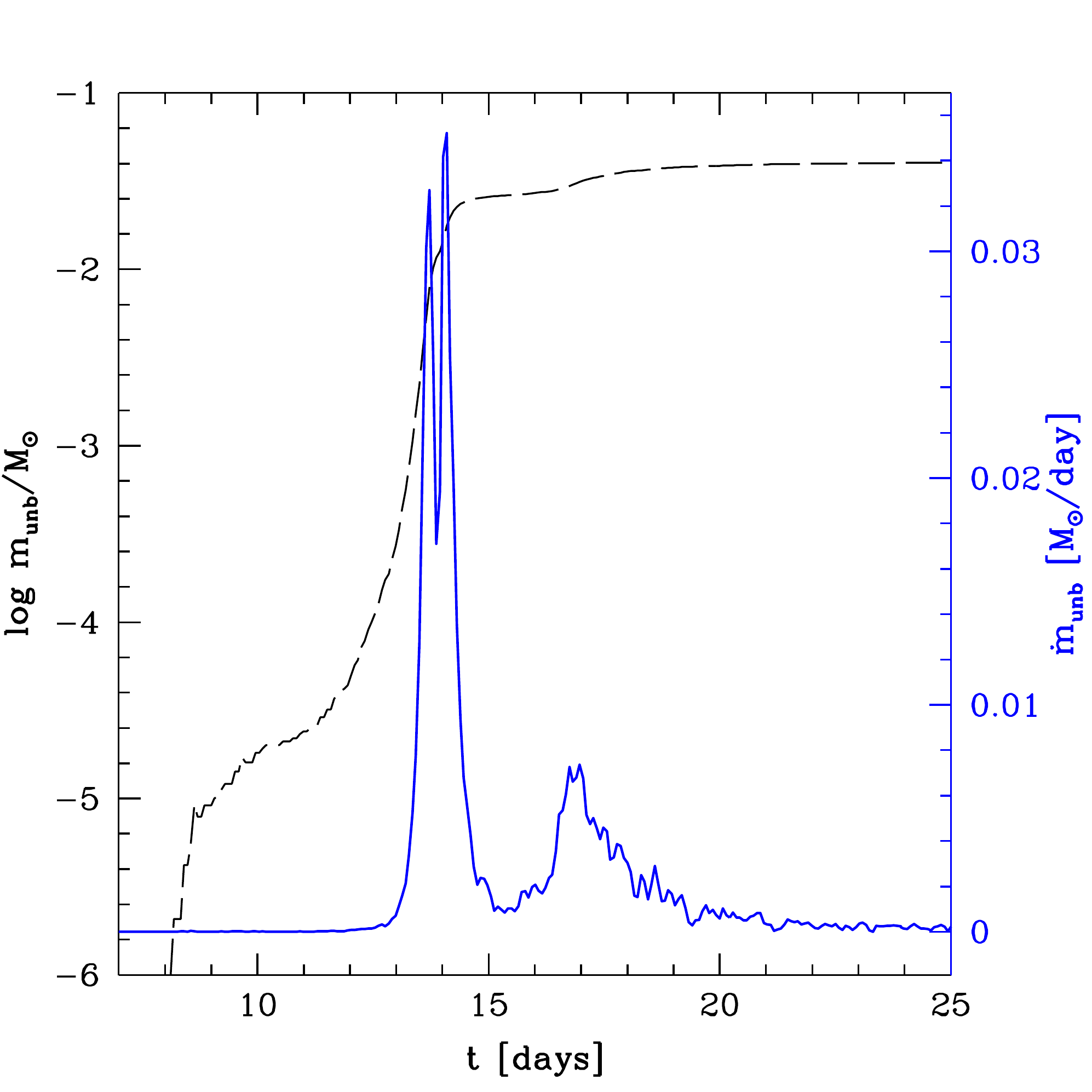}
 \end{center}
\caption{The mass of the ejecta (black dashed line) and its derivative (blue solid line) as functions of time, in the simulation pn351. 
Each peak shown in the plot corresponds to one episode of the mass outburst.}
\label{fig:ejectadot}
\end{figure}

\begin{figure}[t]
 \begin{center}
   \includegraphics[scale=0.45]{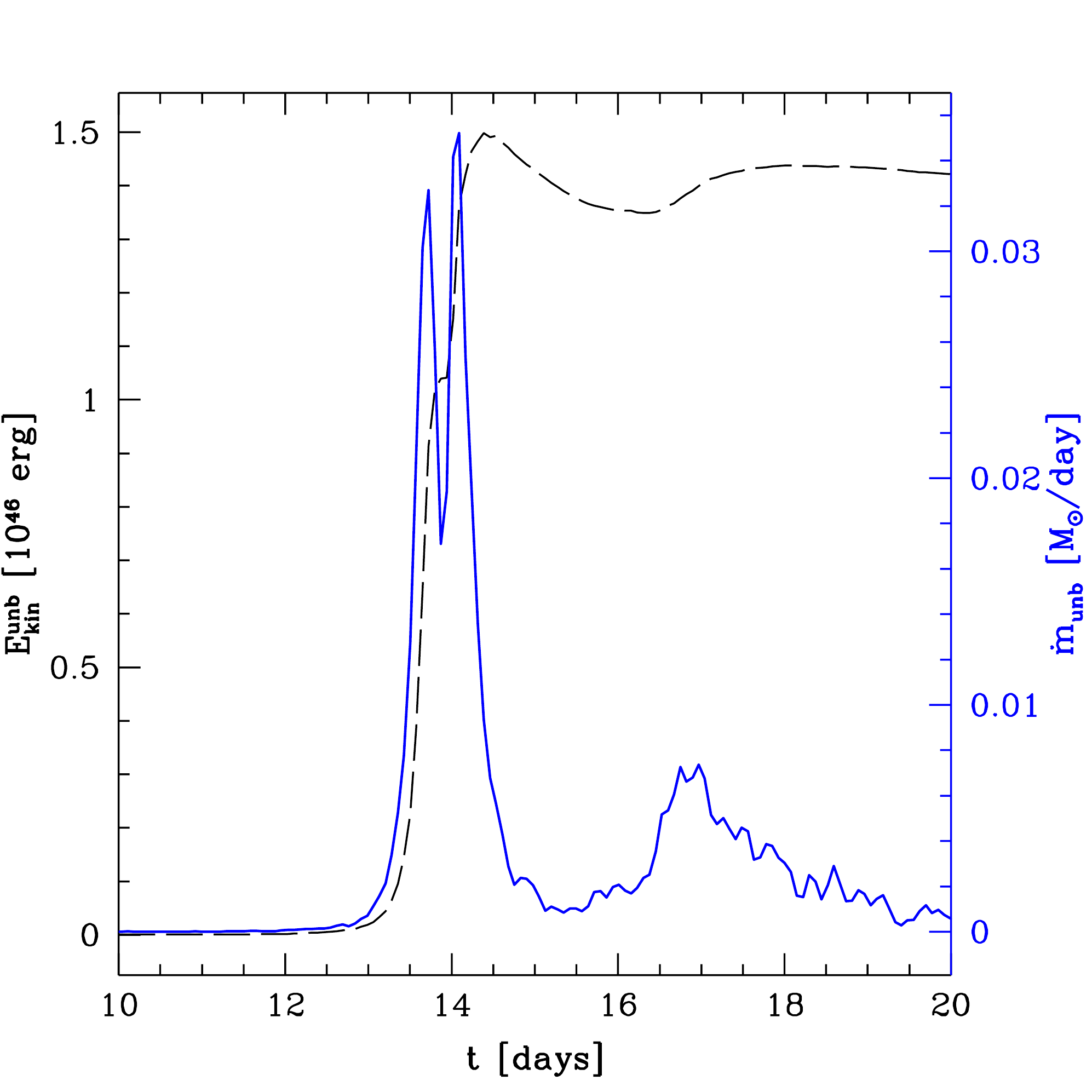}
 \end{center}
\caption{Kinetic energy of the ejecta (black dashed line) and $\dot{m}_{\rm unb}$ (blue solid line) in the simulation pn351.}
\label{fig:kinejectadot}
\end{figure}

We found that the ejection of stellar material usually proceeds 
in several mass outbursts, when the mass loss rate increases significantly for short periods of time and then drops again.
To identify and distinguish these mass-ejection outbursts, 
we compute the change in the mass of the ejecta with time, $\dot{m}_{\rm unb}\equiv dm_{\rm unb}/dt$.
E.g., in  Figure~\ref{fig:ejectadot} we can distinguish three episodes of the mass outbursts, 
each corresponding to a spike in the mass-ejecta rate $\dot{m}_{\rm unb}$.
The first peak corresponds to the mass outburst {\it before} the merger (during the plunge-in), 
the second peak correspond to the mass outburst {\it during} the merger while the last one corresponds to the 
mass outburst {\it after} the merger has been completed. Recall we defined as {\it merger} the time when $a_{\rm orb}<0.1R_\odot$, see \S\ref{subsec:merger}.

A summary of the mass outbursts in different simulations is presented in Table~\ref{tab:permass}. 
We find that the mass-outburst before the merger is absent in synchronized system with a degenerate donor --
the same systems that feature rapid radius expansion and $L_2$ mass loss. 
In contrast, all non-synchronized simulations with a degenerate donor show all three outbursts mentioned above. 
Binaries with a non-degenerate donor do not produce a mass outburst after the merger; instead they always have 2 episodes of mass ejection independent of the initial synchronization.
The duration of all episodes is on the order of the dynamical timescale of the system. 
Most of the ejected material is ejected during the mass-outbursts, 
but not all (see also Table~\ref{tab:tableresvel} for the total mass of the material to infinity).

Compared to the determination of the mass of the ejecta, proper values of the ejecta kinetic 
energy {\it exactly} at the moment when the material was ejected are harder to determine.
This is because the ejection is a continuous process, and each mass outburst can take from a few hours up to several days.
At each time-step we have some particles that are ejected right then, 
but other particles were ejected during the previous time-step, and have already started their travel to infinity 
-- so they already have lost some initial  kinetic energy after overcoming partially the potential well.
An example can be seen in Figure~\ref{fig:kinejectadot} -- the mass of each outburst loses
its kinetic energy with time until it approaches a constant value at infinity.
In Table~\ref{tab:permass} we provide the kinetic energy that all particles contributing to each mass outburst had
at the moment they were identified as unbound for first time.
We anticipate that those values are lower estimates for the kinetic energy of the ejecta,
as velocities inferred from these energies do not greatly exceed the escape velocities. 
However the relative values between the outbursts is more meaningful, and shows that 
ejecta from the initial outburst usually have a higher velocity than those from the second or third outburst.

It was shown that the observed light-curve in V1309~Sco could be reconstructed 
with two mass outbursts of 0.02 and 0.04 $M_\odot$ mass loss, with corresponding 
kinetic energies for each outburst as $0.9\times10^{46}$~ergs and $0.75\times10^{46}$~ergs \citep{2013Sci...339..433I}.
These values are within the range of the obtained values in our simulations (see Table~\ref{tab:permass}),
with non-degenerate and non-synchronized systems being the closest match.

\subsection{Properties of the ejected material at the infinity}

\subsubsection{Velocity of the ejecta}

\begin{table}[t]

\caption{Velocities at infinity.}
\label{tab:tableresvel}
\begin{center}
\begin{tabular}{c l c l l l l}
\hline
 Model & $v_{\rm esc}^{\rm ini}$ & $v_{\rm esc,bin}^{\rm ini}$ &$m_{\rm tot}^{\rm unb}$& $E_{\rm kin,\infty}^{\rm unb} $ & $v^{\rm unb}_{\infty}$ & $j_{\rm unb}/j_{\rm ini}$\\
\hline
 ps334 & 404 & 319 &0.0549 &1.94 & 189 &8.06\\
 mn351 & 396 & 319 &0.0382 &1.51 & 199 &7.74\\
 pn351 & 396 & 319 &0.0415 &1.30 & 177 &7.50\\
 ps351 & 398 & 319 &0.0800 &2.28 & 169 &6.96\\
 ms376 & 383 & 313 &0.0470 &1.46 & 177 &8.01\\
 ps376 & 383 & 313 &0.0466 &1.75 & 194 &8.08\\
 ms372 & 382 & 317 &0.0464 &1.55 & 183 &8.10\\
 ps379 & 381 & 319 &0.0808 &2.26 & 168 &5.88\\
 pn344 & 400 & 317 &0.0415 &1.33 & 180 &7.86\\
 ps375 & 383 & 317 &0.0859 &2.40 & 168 &6.16\\
 mn344 & 400 & 317 &0.0362 &1.21 & 183 &8.01\\
 ms375 & 383 & 317 &0.0479 &1.65 & 186 &7.92\\
 pn319 & 415 & 317 &0.0583 &1.80 & 176 &7.88\\
\hline
\end{tabular}
\begin{tablenotes}
       $v_{\rm esc}^{\rm ini}$ is the escape velocity from the surface of the initial primary, $v_{\rm esc,bin}^{\rm ini}$ is the escape velocity from the initial binary (using $a$), $m_{\rm tot}^{\rm unb}$ is the total unbound mass in $M_\odot$, $E_{\rm kin,\infty}^{\rm unb}$  is kinetic energy at infinity in $10^{46}$ ergs, $v^{\rm unb}_{\infty}$ is the velocity of the ejecta at infinity, $j_{\rm unb}/j_{\rm ini}$ is the ratio between the specific unbound angular momentum and specific initial angular momentum. All velocities are in km/s.
     \end{tablenotes}
\end{center}
\end{table}

\begin{figure}[t]
 \begin{center}
   \includegraphics[scale=0.45]{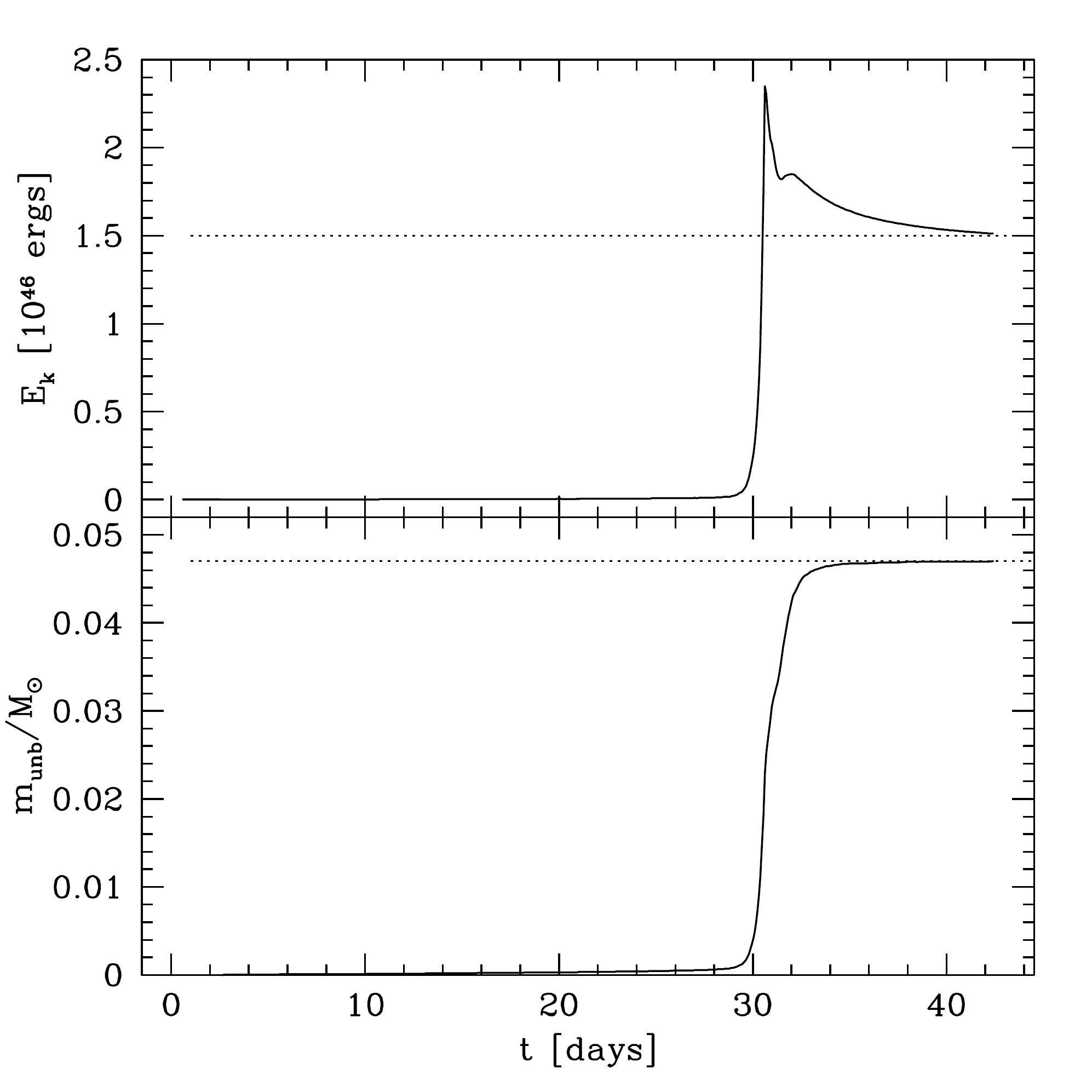}
 \end{center}
\caption{The kinetic energy (top panel) and mass (bottom panel) of the unbound material in the simulation ms376.
The dotted lines indicate the final values at infinity.
}
\label{fig:asymptms376}
\end{figure}

The energy formalism commonly used to evaluate an outcome of a CE event assumes that the material 
is ejected with energy just sufficient to move that material to infinity and that its kinetic energy there is zero and hence does not need to be taken into account in the energy balance.
We find that the kinetic energy of the ejected material is significantly non-negligible at infinity, 
moreover, it is not much different than at the moment it was just ejected (see Figure~\ref{fig:asymptms376}). Since at the onset of the simulations the two stars are considered point masses, we can compute the initial orbital energy by simply using $E_{\rm orb}=GM_1M_2/(2a_{\rm orb})$, where $M_1=1.52M_\odot$ and $M_2=0.16M_\odot$, and $a_{\rm orb}\simeq6.3R_\odot$ (see table \ref{tab:tablesol}). Hence, $E_{\rm orb}\sim 7.3\times 10^{46}$ erg which can be compared with the kinetic energy given by the table \ref{tab:tableresvel}. We can conclude that the ejecta takes away up to 1/3 of the initial orbital energy. \revone{Note that we did not use the CE energy formalism anywhere in our calculations.}

In Table \ref{tab:tableresvel} we show the velocities of the ejected material at infinity, $v_{\infty}^{\rm unb}$, 
and they are as large as 42\% to 51\% of the initial escape velocity from the surface of a donor.
The velocities we obtain are well consistent with the average velocities of the ejecta from the observations, 
160-180~km~s$^{-1}$ \citep{2010A&A...516A.108M}. We note that, as with the velocity at the moment of the ejection,
at infinity there is also no single-valued velocity for all the ejected material, 
and ejecta speeds are usually significantly higher at the start of the mass loss, for the outer layers, and
smaller for material ejected after the merger is complete.

\subsubsection{Angular momentum}

The total angular momentum carried away by the ejecta is between 17\% and 33\% 
of the initial total angular momentum of the binary, even though it is taken away by an extremely small amount of the material
(see Table~\ref{tab:tableang}).
The specific angular momentum of the ejected material exceeds the initial specific angular momentum by a factor of 5.8 to 8.1 (see Table~\ref{tab:tableresvel}).
The relative fraction of the total angular momentum that is carried away with the ejecta is highest in the simulations
with a synchronized binary and a degenerate donor and smallest in non-synchronized binaries with a non-degenerate donor.

\subsubsection{The entropy and temperature of the ejecta}

\begin{figure}[t]
 \begin{center}
   \includegraphics[scale=0.45]{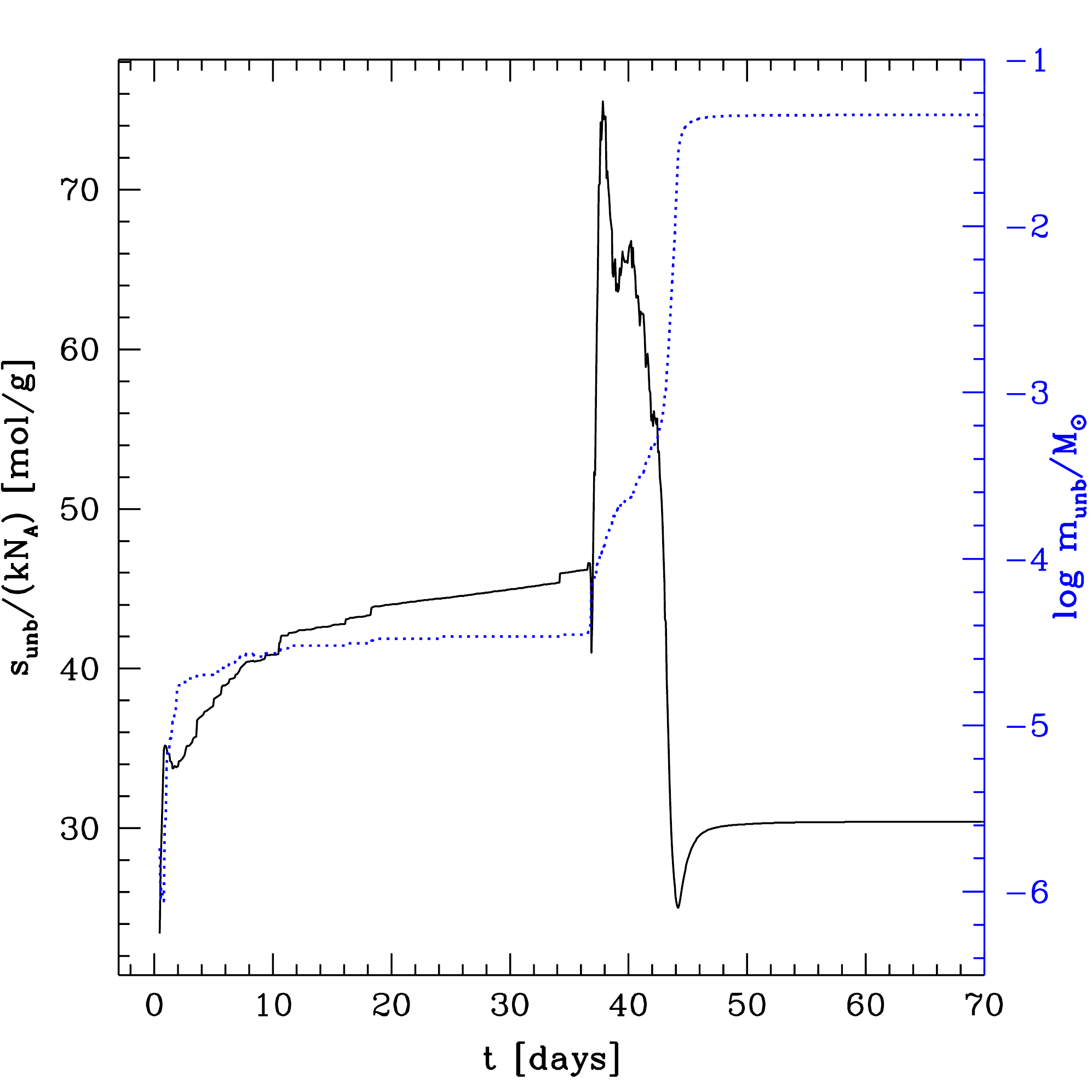}
 \end{center}
\caption{Evolution of the specific entropy (black solid line) and mass (blue dotted line) of the ejecta in simulation ps376.}
\label{fig:unbentropy}
\end{figure}

\begin{figure}[t]
 \begin{center}
   \includegraphics[scale=0.45]{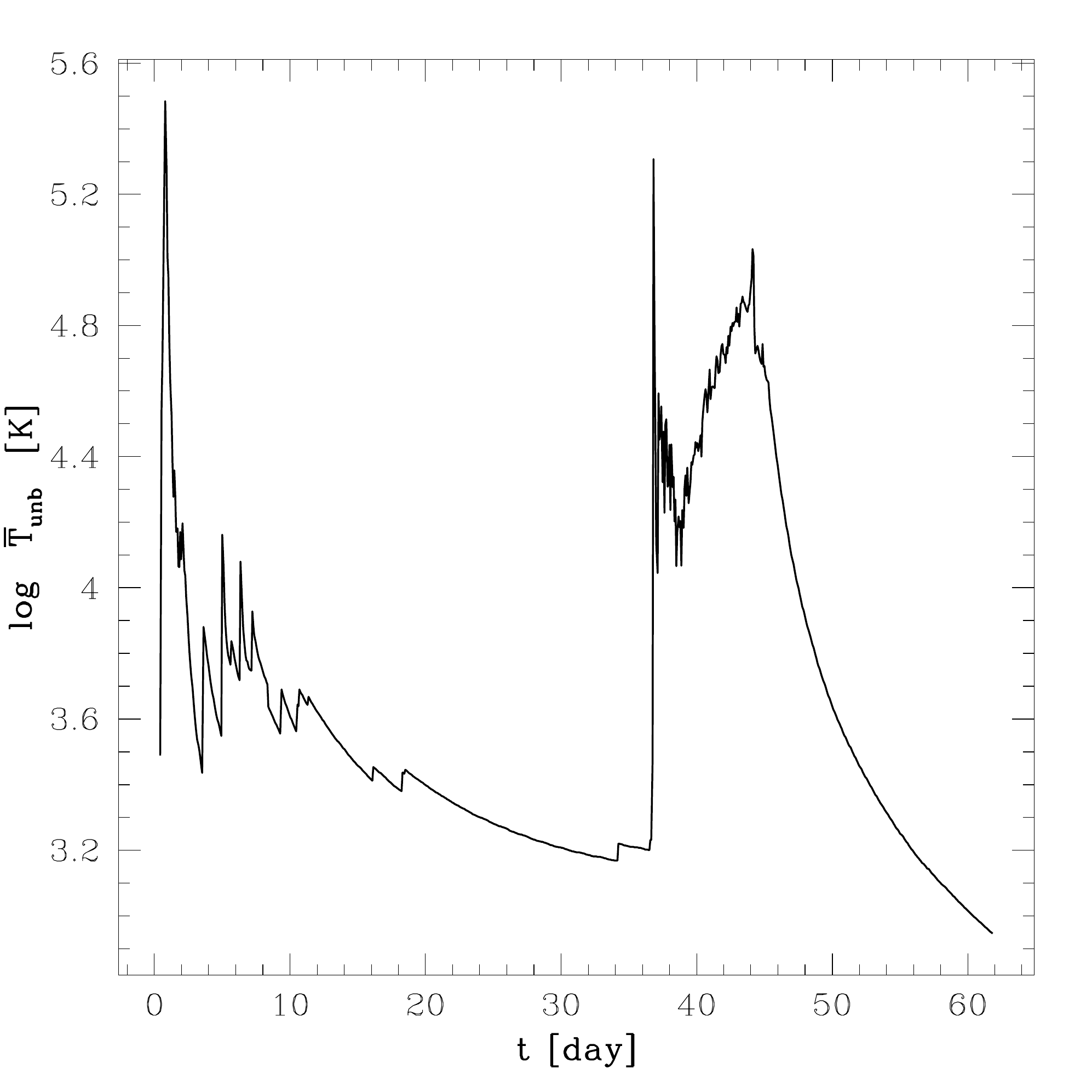}
 \end{center}
\caption{Average temperature of the unbound material (simulation ps376). }
\label{fig:Tvst85258}
\end{figure}

The specific entropy of the material in the envelope of the unperturbed donor, $s/(kN_{\rm A})$, is about 22 mol/g.
The initial mass loss starts when the common envelope has not yet formed, but the ejected
material is already shock-heated, with its specific entropy increased by about 20 mol/g  (see Figure \ref{fig:unbentropy}). 
Once the common envelope forms and the companion starts its spiral-in, the ejected 
material is the most shock-heated throughout the complete event -- its specific entropy exceeds its initial value by up to 50 mol/g. 
As the companion continues to  plunge-in, more of the envelope of the donor gets ejected, 
but this material is already less shock-heated, and overall the entropy of the ejecta decreases and 
reaches a minimum -- the big dip that takes place at about $t_{\rm merg}$.
As the merger is completed and the ejected material evolves adiabatically, its entropy remains constant,  
at a value about 8 mol/g  higher than the initial value of the specific entropy in the donor.
This general behavior is characteristic for all the simulations, while the final and maximum entropy values varying somewhat from case to case.

\begin{figure*}[t]
 \begin{center}
   \includegraphics[scale=0.9,angle=-90]{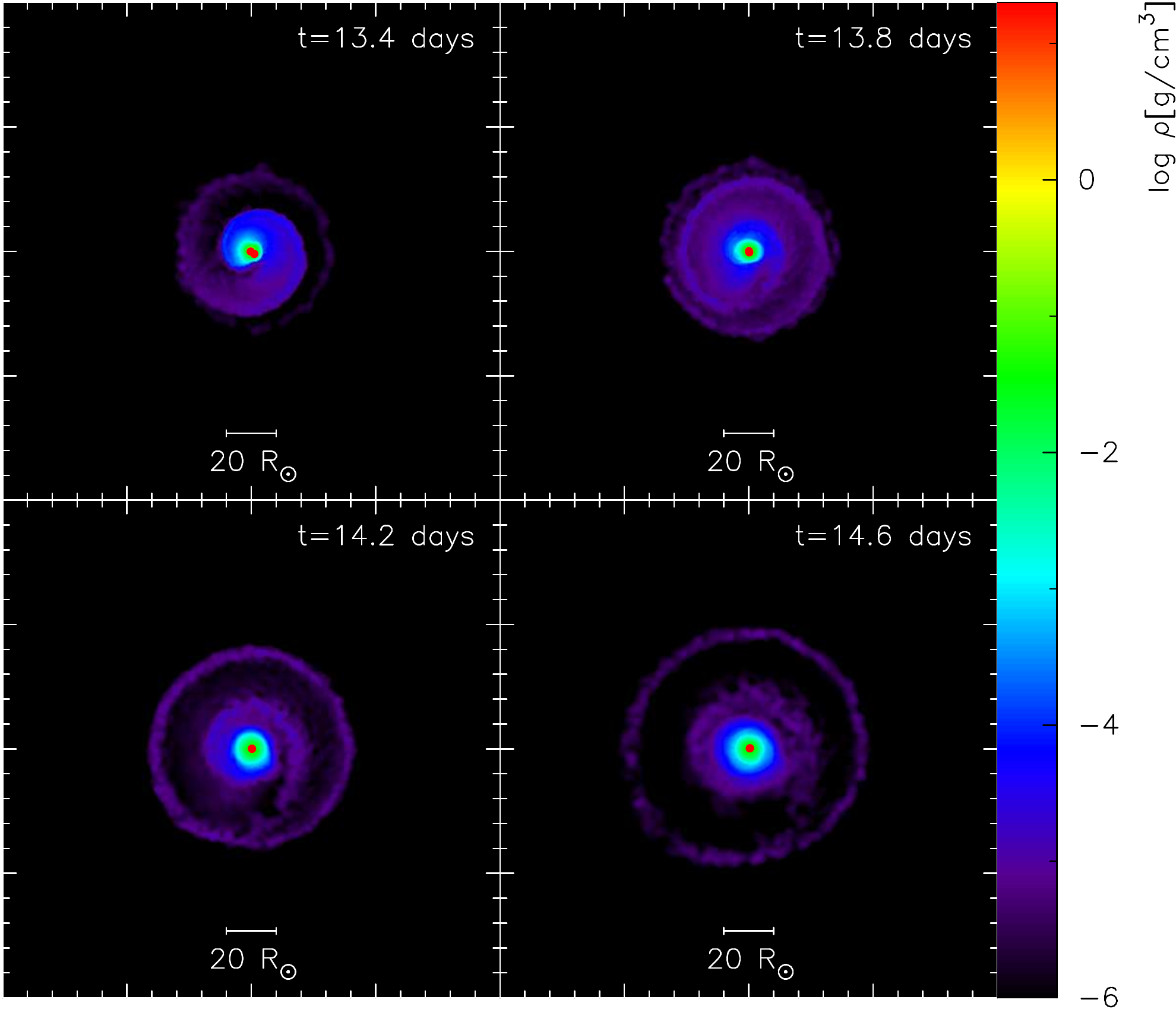}
 \end{center}
\caption{Ring formation in the simulation ps375.}
\label{fig:ringps375}
\end{figure*}

Figure \ref{fig:Tvst85258} shows the evolution of the average temperature for the unbound material, in the same 
simulation ps376 as for the specific entropy discussed above. 
The first spike in the simulation takes place at the first contact between the low-mass companion and the surface of the giant
and involves only a small quantity of ejected material.
The second spike corresponds to the first episode of the mass ejection -- when the common envelope gets formed.
The third spike is due to the shock-heating during the merger. 
After the merger is completed and there are no more mass outbursts, the ejecta temperature 
demonstrates adiabatic cooling.
In our simulations, the equation of state does not include ionization. In a real merger, 
the recombination process will undoubtedly change the temperature evolution \citep{2010ApJ...714..155K,2013Sci...339..433I}.

\subsection{Ejecta appearance}

The ejection proceeds in mass outbursts. If outbursts are well separated by 
a minimum in the rate of the mass loss, they might be distinguished  
even when the ejecta is very far from the merged object.
Indeed, in all our simulations  with a synchronized binary, the ejecta appears in the form of a ring. 
In Figure~\ref{fig:ringps375} we show the formation of a typical ring structure from the initially spiral-shaped outflow. 
In the case shown, the outer ring is formed by the material
from  the mass outburst during the merger, and the inner, less pronounced ring, by the last mass outburst.
We do not see a ring or another well distinguished structure formation in simulations with 
a non-synchronized binary -- there, the ejecta is rather isotropic with many ``clumps''.

\section{Merger product}

The observations of V1309~Sco  after  the outburst, when its light-curve was rapidly declining,
show that its temperature is cooler than that of its progenitor. 
Its radius, inferred from luminosity and temperature, reached  $\sim 310 R_\odot$ at maximum.
In about 15 days after the peak, the radius of the object was estimated to have reduced already to
$\sim 150\ R_\odot$, and in a few years it shrunk to about $5 R_\odot$, 
just a bit larger than the progenitor \citep{2011A&A...528A.114T}.
While the implied radius of $ \sim 150  R_\odot$ during the luminosity plateau is related to 
the wavefront of cooling and recombination of the ejecta \citep{2013Sci...339..433I}, the observations during the light-curve decline
correspond to the surface of the merged star that becomes visible 
once the ejecta has fully recombined and become transparent.

\subsection{Equilibrium}

\begin{figure}[t]
 \begin{center}
   \includegraphics[scale=0.37]{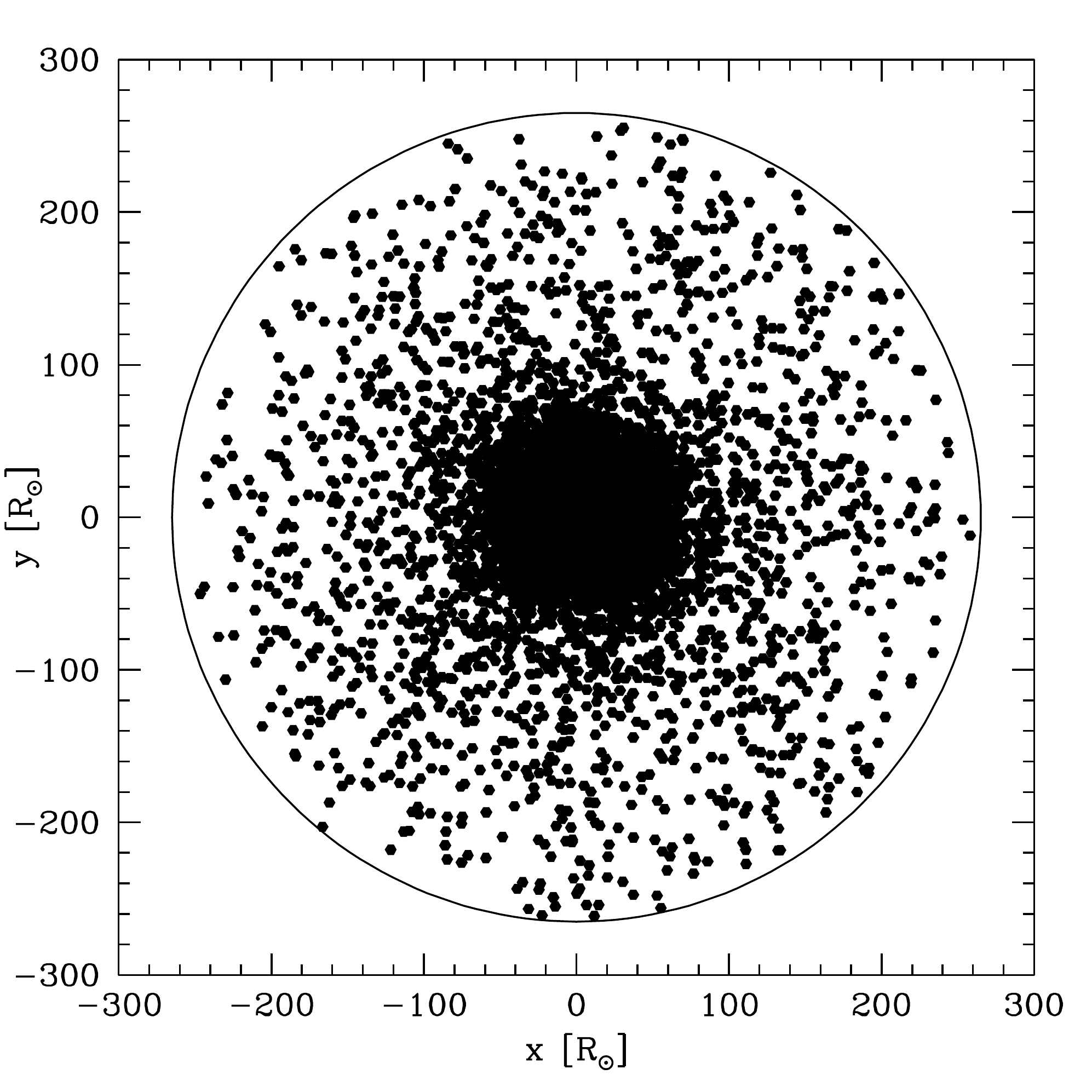}
 \end{center}
\caption{Distribution of particles projected onto the equatorial plane when the merger product has reached its hydrostatic equilibrium for the simulation ms376.}
\label{fig:eq_starms376}
\end{figure}

\begin{figure}[t]
 \begin{center}
   \includegraphics[scale=0.37]{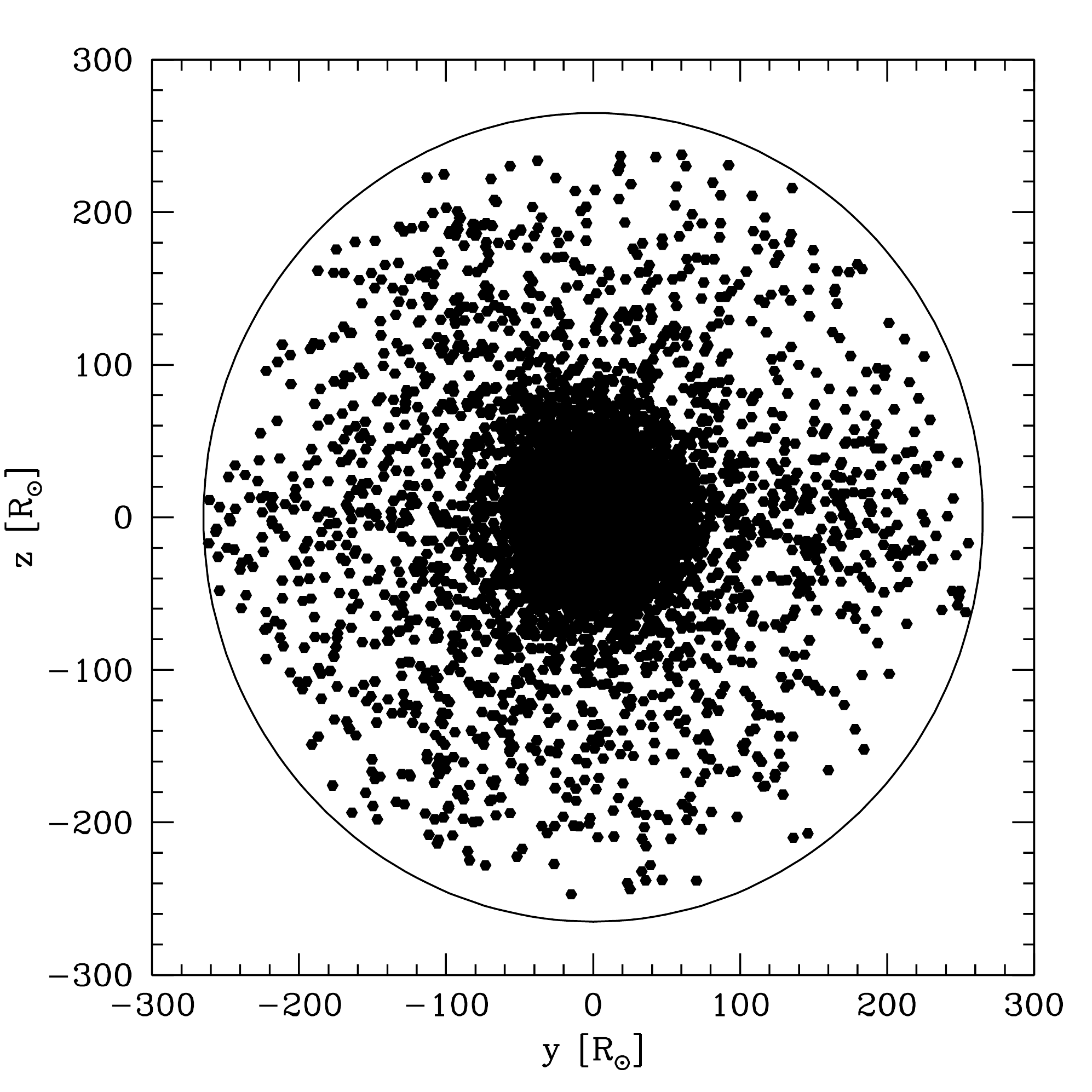}
 \end{center}
\caption{Distribution of particles projected onto the $yz$ plane when the merger product has reached its hydrostatic equilibrium for the simulation ms376.}
\label{fig:sym_starms376}
\end{figure}

We analyze the merger product once it is in hydrostatic equilibrium. 
We define this  as when the kinetic energy of the merged object is much smaller than 
its internal energy, $E_{\rm in}/E_{\rm kin}>30$, and the internal and kinetic energies 
of the merger product remain nearly constant for a time interval comparable 
to its dynamical timescale. By that moment  at least $1.6 M_\odot$ of the bound mass 
is enclosed in a radius less than $100\ R_\odot$ 
(see Figures \ref{fig:eq_starms376} and \ref{fig:sym_starms376}). 
The same radius of $100\ R_\odot$ corresponds roughly to the surface of optically thick material. 
More specifically, in simulations with a non-degenerate companion, the bound objects are fully
enclosed in a radius less than $120\ R_\odot$, while in simulations with a degenerate companion
bound material can extend much farther away, up to $\sim 600 R_\odot$.
The dynamical timescale for a $1.6 M_\odot$ object of  $100 R_\odot$  
is about 15 days, and, as we found from the simulations, the kinetic and thermal energies usually 
stabilize in about half that time. 
The low-mass expanded envelope can be expected to contract on its own thermal timescale, 
which is just about a few years, as in observations. 
This thermal contraction phase, with the rapid loss of energy 
from the envelope with radiation, however, can not be modeled with the SPH code, 
despite the timescale being close to the dynamical timescale.

\subsection{Symmetries}

We find that the distribution of particles of the merger product is fairly symmetric, both in the equatorial ($xy$) plane
and the polar axis ($yz$) plane; however, rotation flattens the merger product so that there are more particles near the equator than near the 
poles (see Figures~\ref{fig:eq_starms376} and Figure~\ref{fig:sym_starms376}).

We calculate the mass of the bound material in different directions. The Northern and Southern hemispheres 
have a very similar mass (except for the core), with the ratio of masses very close to one. 
Similarly, we also calculate the mass enclosed in a cone with an opening angle of $50^\circ$ along $+x$, $+y$ and $+z$ directions. 
These numbers reveal that the ratio between $m_{+x}$ and $m_{+y}$ is about one, while the mass ratios 
of $m_{+x}/m_{+z}$, and $m_{+y}/m_{+z}$ are up to 1.5 in the case of a non-synchronized binary with a main-sequence donor.
At the same time, a synchronized binary with a main-sequence donor is almost symmetric.

\begin{figure}[t]
 \begin{center}
   \includegraphics[scale=0.37]{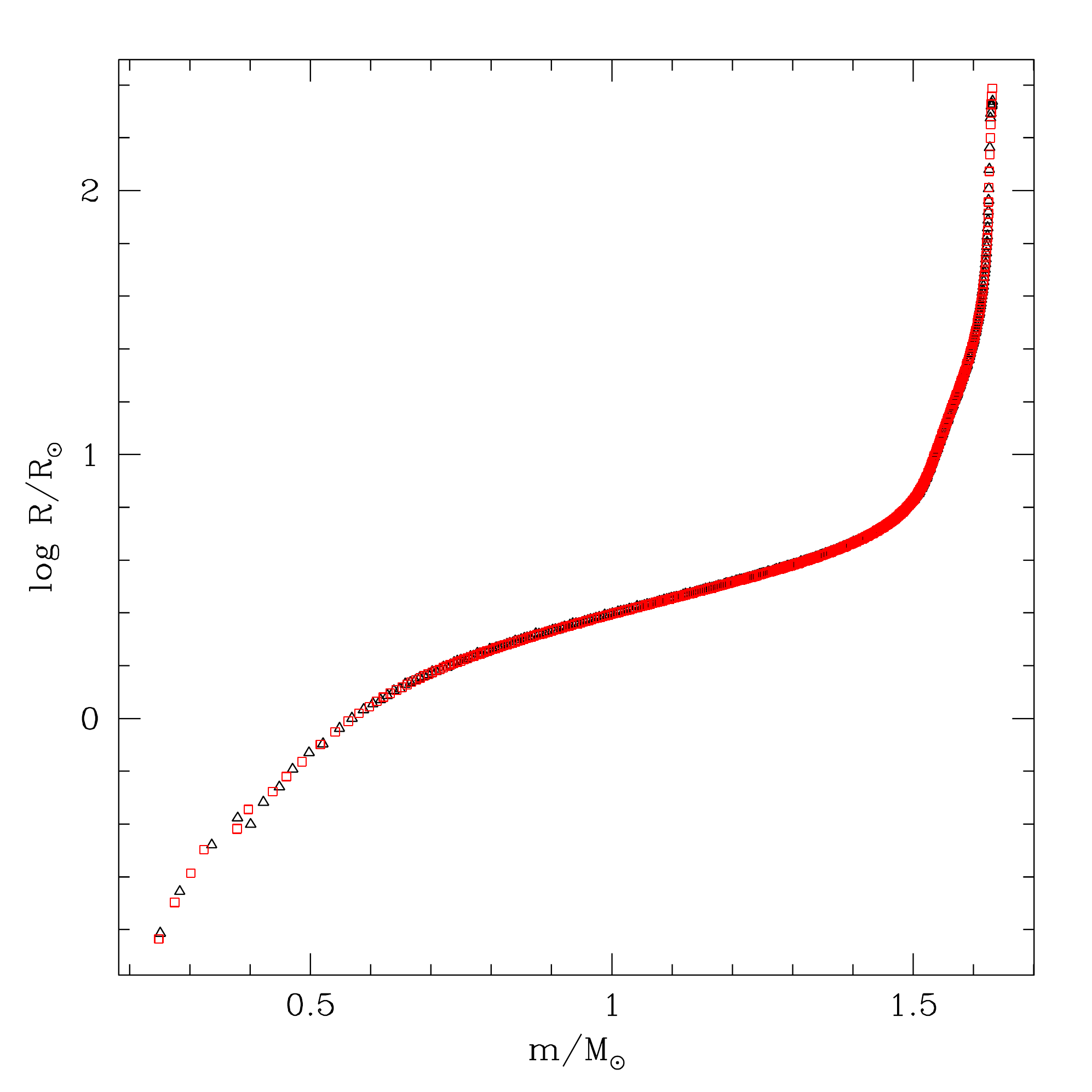}
 \end{center}
\caption{Radius profiles for the merger product for the simulation ms376. The black open triangles show the profile for the merger product sorted by pressure while the red open boxes are sorted by radius. See the text for details about the sorting methods.}
\label{fig:profradius}
\end{figure}

\subsection{Star profiles}

For rotating stars, it is argued that the stellar equations should be solved across  isobaric shells instead of spherical shells \citep[e.g.,][]{2000ApJ...528..368H}. Accordingly, the transformation from a 3D SPH model to a 1D model can be done by averaging on isobaric surfaces.

We sorted the particles by means of two methods, (i) pressure, and (ii) position/radius; where the particles with maximum pressure is defined as the center in the method (i), and the center of mass of the bound material is defined as the center of the merged product for the method (ii). Once we have sorted the particles by pressure or position/radius, we average the thermodynamics and dynamical variables of the SPH particles by a regular fixed bin.

\begin{figure}[t]
 \begin{center}
   \includegraphics[scale=0.37]{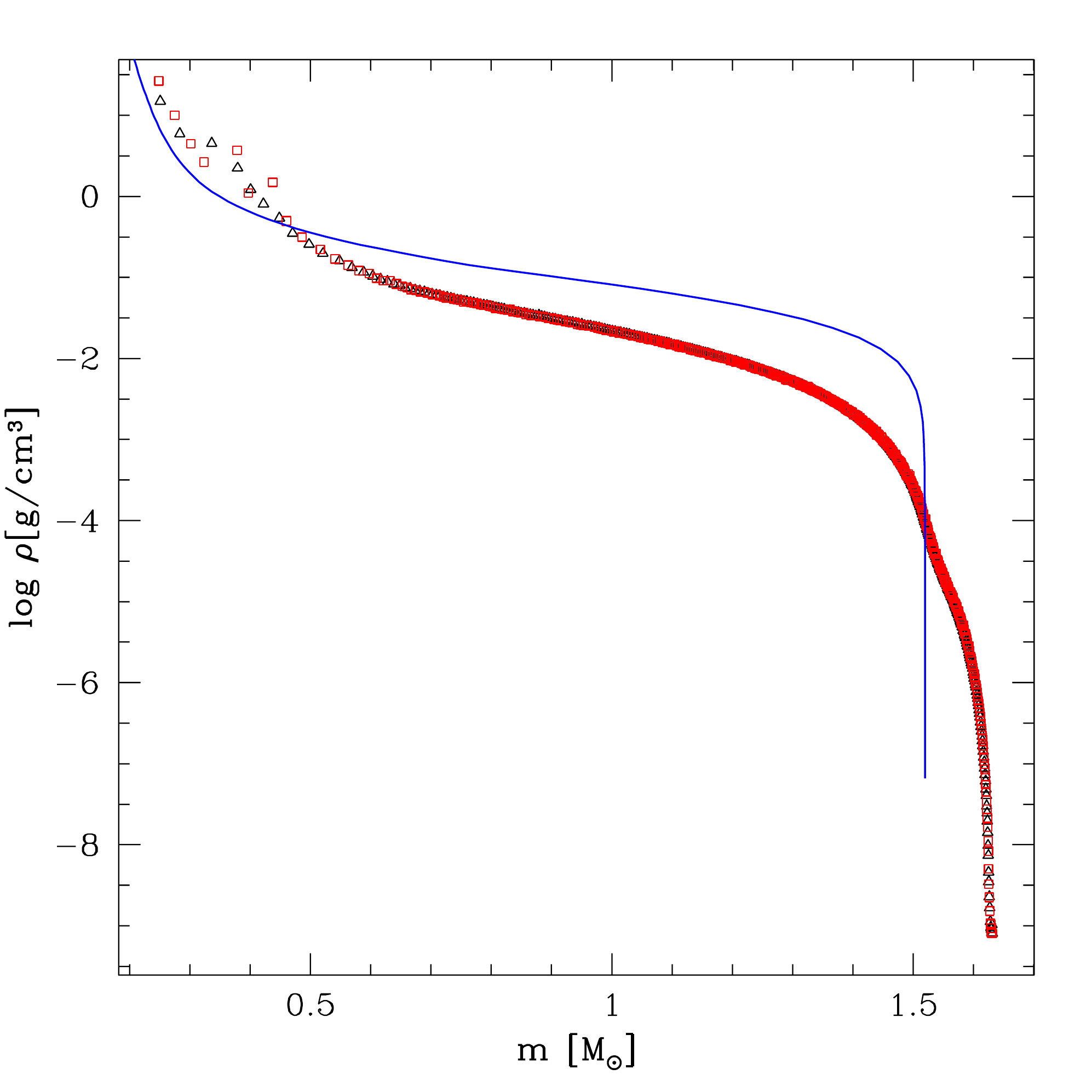}
 \end{center}
\caption{Density profiles for the merger product for the simulation ms376. The black open triangles show the profile for the merger product sorted by pressure, while the red open boxes are sorted by radius. The blue solid line corresponds to the initial profile of the RG.}
\label{fig:profdensity}
\end{figure}

\begin{figure}[t]
 \begin{center}
   \includegraphics[scale=0.37]{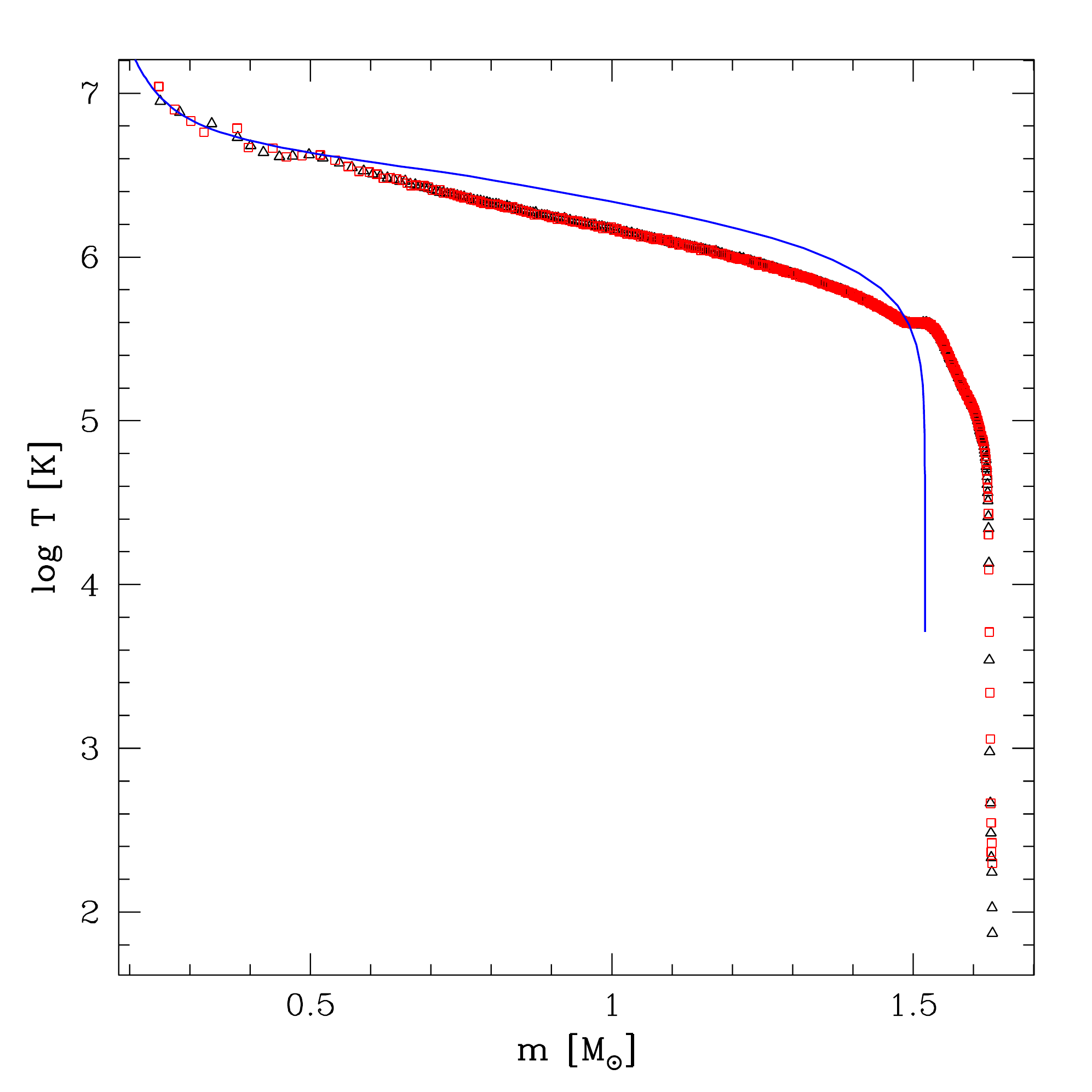}
 \end{center}
\caption{Temperature profiles for the merger product for the simulation ms376. The black open triangles show the profile for the merger product sorted by pressure, while the red open boxes are sorted by radius. The blue solid line corresponds to the initial profile of the RG.}
\label{fig:proftemperature}
\end{figure}

In Figure~\ref{fig:profradius} we compare  the radius profiles of the formed star (the model ms376) obtained by
the radius-sorting and pressure-sorting methods. Except for the very inner part near the core
the radius profiles obtained with two methods are indistinguishable. 
A similar comparison of density and temperature profiles obtained with the two methods show
that there is a slight effect for the temperature near the surface (it is lower when isobaric surfaces are used), 
but otherwise both methods continue to give similar results (see Figures \ref{fig:profdensity} and \ref{fig:proftemperature}).

\subsection{Angular momentum}

\begin{figure}[t]
 \begin{center}
   \includegraphics[scale=0.37]{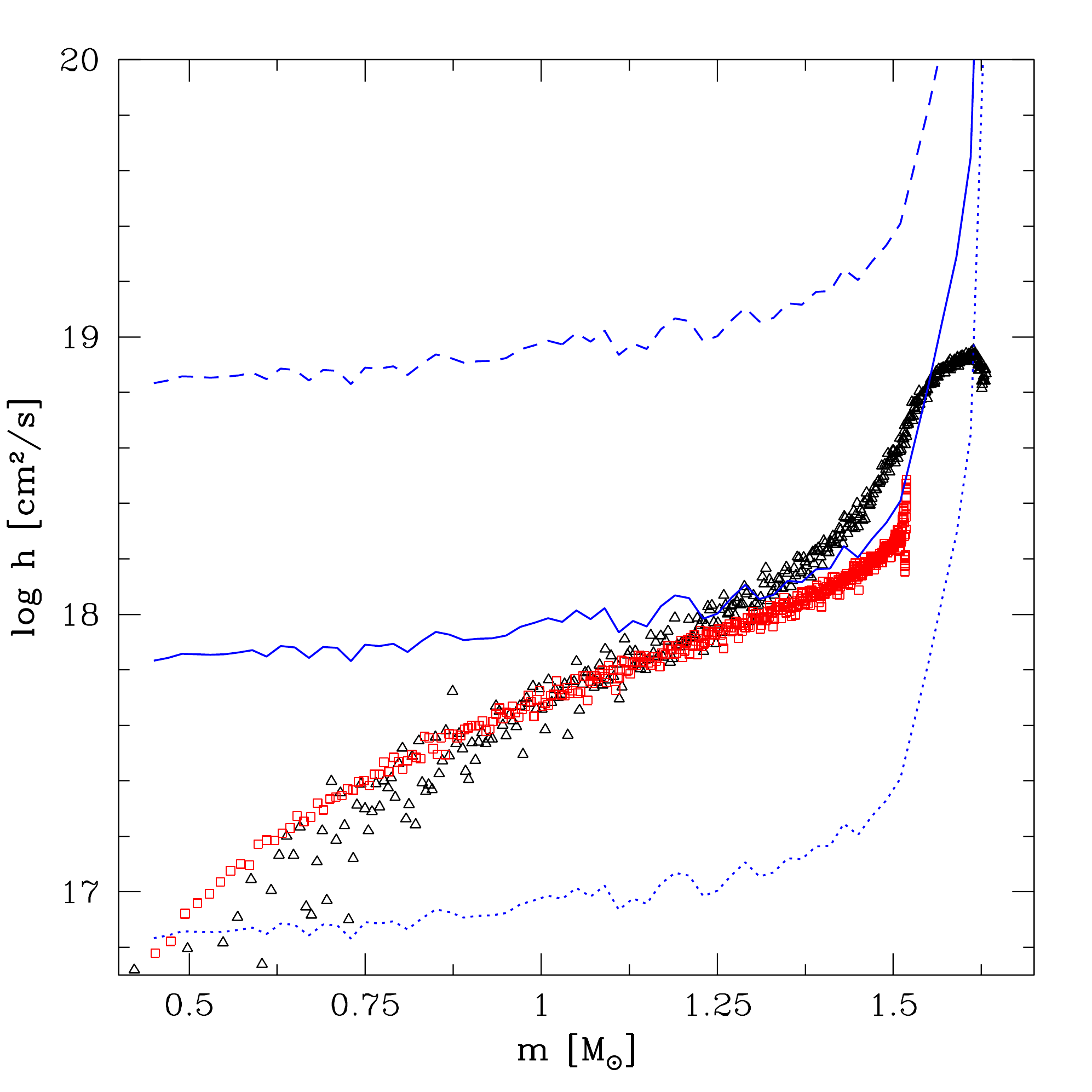}
 \end{center}
\caption{Specific angular momentum profile for the merger product (black open triangles) and the primary star (red open boxes), for the simulation ms376. The blue lines show the specific angular momentum for $\Omega=10^{-4}$ Hz (dashed line), $\Omega=10^{-5}$ Hz (solid line) and $\Omega=10^{-6}$ Hz (dotted line). }
\label{fig:angmomprod}
\end{figure}

We find that the merger product does not rotate as a solid body -- 
see Figure~\ref{fig:angmomprod}, where we show the specific angular momentum profile of the merger product in simulation ps376 compared to that of the initial RG star and that of the merger product if it were to rotate rigidly. 
We note that our simulations use the Balsara switch in the artificial viscosity, in order to minimize the spurious transport of angular momentum \citep{1999JCoPh.152..687L}.

The envelope of the merger product, from about 1.25 $M_\odot$ to 1.5 $M_\odot$, shows rotation close to that of a rigid body
with $\Omega=10^{-5}$~Hz -- note that the star is still expanded, which is why the rotation 
appears to be slower than in the initial star. 
Overall, the merger product has 2-3 times more angular momentum than in the initial star.
If the merger product contracts to $5 R_\odot$ within several years, 
as in the observations, the critical surface angular velocity will become $\Omega_{\rm crit}=10^{-4}$~Hz and the critical value 
of the specific angular momenta near the surface will be  
$j_{\rm crit}\sim10^{19}$ cm$^2$~s$^{-1}$. It is possible therefore (although not necessary) 
that the outer layers of the merger product might rotate close to the critical rate.
We also note that the rotational profile with $dh/dr <0$ is secularly unstable 
on the thermal timescale of the star \citep{1969A&A.....2..309K}.

\subsection{Entropy}

\begin{figure}[t]
 \begin{center}
   \includegraphics[scale=0.37]{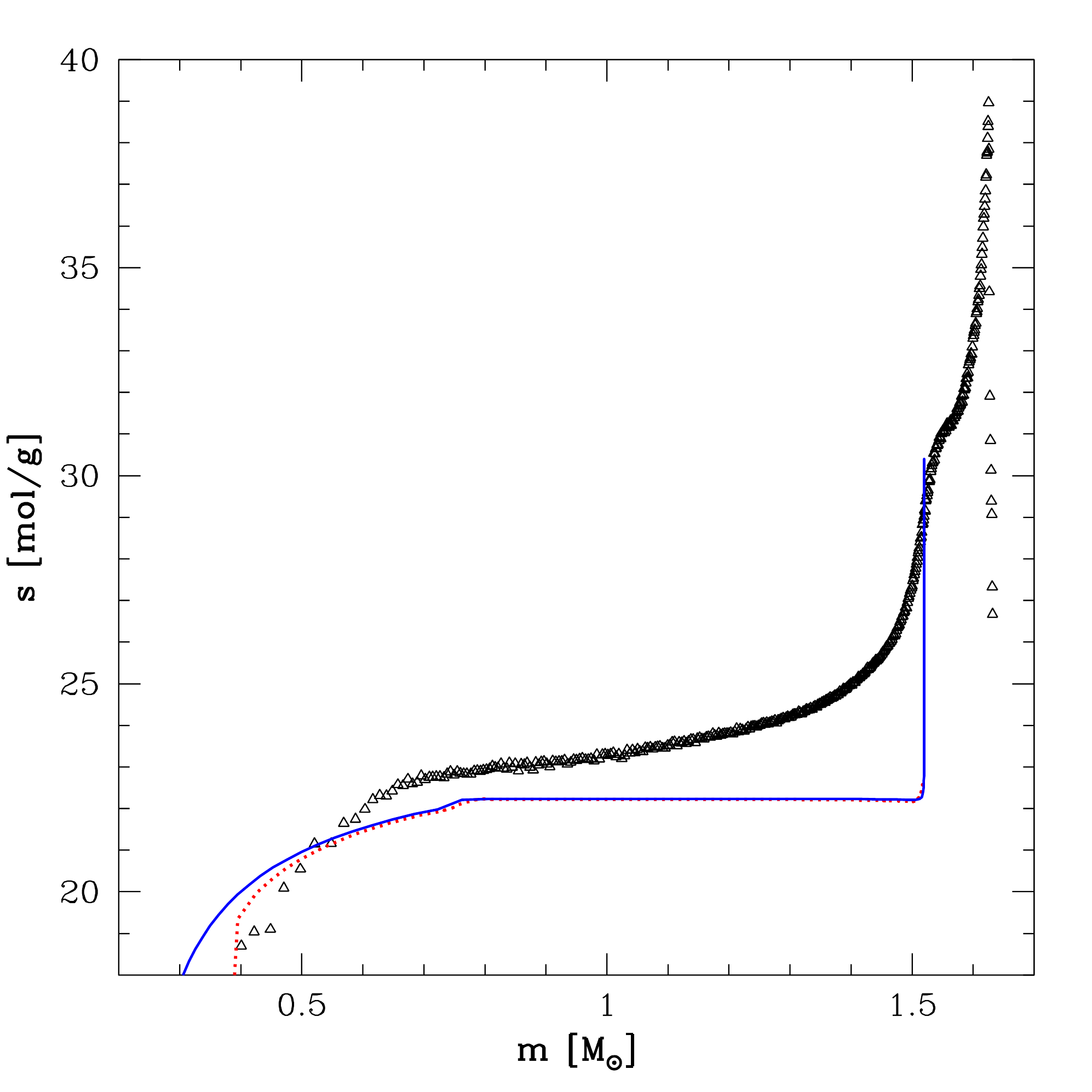}
 \end{center}
\caption{Specific entropy profiles for the merger product for the simulation ms376 assuming a fully ionized gas. The black open triangles show the profile for the merger product, the red dotted line corresponds to the profile for the relaxed primary in SPH, while the blue solid line corresponds to the profile given by the stellar model before the relaxation in SPH.}
\label{fig:specentroprod}
\end{figure}

In Figure \ref{fig:specentroprod} we show the specific entropy of the merger product, comparing it to the initial entropy profile of the donor as given both by the 1D stellar evolution code and  by the relaxed 3D SPH model. 
All entropies here are obtained using Equation~\eqref{eq:totalentropyfully} that takes into account chemical composition and radiation. This equation, however, neglects partial ionization. This results in the artificial peak near the surface in the entropy profile of the initial stellar model. Otherwise, the relaxed 3D star and the initial 1D star are very similar everywhere except very close to the core, which is represented by an artificial particle. The merger product has been shock-heated throughout, with no trace of either the convective envelope with uniform specific entropy of about 22 mol/g or of its convective companion that had a uniform specific entropy of about 12 mol/g.

\section{Discussion}

In this paper we have studied the V1309~Sco outburst by
adopting a model in which the outburst results from the merger of a binary consisting of a
$1.52 M_\odot$ giant and a $0.16 M_\odot$ companion, with a 1.44 day orbital period prior to merger.
We have analyzed how initial conditions such as the nature of the companion (whether it was a white dwarf or a main sequence star)
and the initial synchronization between the orbit and the rotation of the donor
could affect the dynamical evolution before, during, and after the merger,
and we have compared our results to available observations.

For this analysis we have developed a set of tools that allow us to quantify the numerical 
simulations of common envelope events in a general case, independently of whether the 
event under consideration would result in a merger, as in the case of V1309~Sco, or in a binary formation.
In presenting this set of tools, we have specifically discussed:

\begin{itemize}
\item how to compute the effective radius of the donor star -- we have found that the volume equivalent radius is best
when considering mass transfer, \S2.2, also \S3.3;
\item how to define the orbital period in a binary decaying into a CE -- we have analyzed the binary from the point of view of
instantaneous (found from Keplerian orbits)  and apparent (visually detectable) orbital periods, \S3.1;
\item the relevant characteristic timescales  -- we have discussed how to determine quantitatively  
 the start of the CE in simulations, as well as the start of the plunge-in and the moment of the merger, \S3.5;
\item how to distinguish the unbound material -- we have determined that
the positiveness of the sum of only the potential and the kinetic energies for an SPH particle should be used as a criterion for its unboundedness and that inclusion of the internal energy in the criterion can lead to an error in the 
identification of the ejected material, \S3.2 and \S5.1. 
\item the symmetry of the post-common envelope merger product -- we have compared the degree of asymmetry of the formed star in different directions 
as well as compared the mapping of the 3D structure to a 1D profile using spherical and isobaric surfaces, \S6.2 and \S6.3. 
\end{itemize}

By comparing the initial orbital angular momenta to the critical angular momentum determined from the spin of the stars, we have showed that 
 all the binary configurations we have considered should be affected by the Darwin instability 
at the start of the simulations.
But what can be expected from the Darwin instability, what are the timescales for the orbital-decay evolution that we obtain,  and 
how reasonable are these timescales when compared against observations?
 
As expected, a longer timescale for the orbital decay takes place 
in a synchronized binary, as compared to a non-synchronized binary with all other initial conditions being similar.
Our longest simulation proceeds for $\sim 150$~days before the merger. During 
the {\it early-stage} orbital decay (before any mass transfer), the orbital-decay timescale $P/\dot P$ is as long as several decades, while, by the end, the decay has accelerated significantly and is much faster.
We also are able to fit some of our simulations with the observationally obtained exponential decay; the latter  
implies the match for $\dot P/\ddot P$. This suggest that the timescales in the simulations and in the observations
are similar and are on the order of several to a dozen years before significant mass loss.

Note that there is no comparison between the merger timescale from the simulation and the observational ones, as we cannot know the value of $R/R_{\rm rlof}$ at the start of the observations. The initial value of $R/R_{\rm rlof}$ is very important, as the decay timescale depends sensitively on how close the donor is to its RLOF. A donor that is 99\% to its RLOF would merger easily 1000 times faster than a donor that is 94\% to its RLOF, even if observationally they would have the same orbital periods. For our characteristic timescales $t_0-t_{\rm merg}$, the numbers are not terribly different than the observations, as we have $t_0$ being larger than the merger time by up to several hundred days, similar to in the observational fitting from \cite{2011A&A...528A.114T}, where $t_0=2 455 233.5$, and, if we assume that $t_{\rm merg}$ is roughly $2 454 530$, then $t_0-t_{\rm merg}$ is about 700 days.

But can the Darwin instability itself provide such a fast dissipation of the orbit, or could another reason
be primarily responsible for the orbital evolution?
\cite{2011A&A...528A.114T} discussed that, along with the Darwin instability, 
it is possible that 
the merger could have started because the system entered into deep contact, a scenario contemplated by 
\cite{1976ApJ...209..829W,1977ApJ...211..881W}, and started to lose mass via $L_2$.
It is expected that the Darwin instability would act on the  timescale of tidal friction, $\tau_{\rm TF}$, and 
that $\dot P/\ddot P\sim \tau_{\rm TF}$\citep{2001ApJ...562.1012E}, while the $L_2$ mass loss  
is expected to act on the same timescale as $L_1$ mass transfer \cite{1976ApJ...209..829W,1977ApJ...211..881W}.
\cite{2013arXiv1307.4088P} has  argued that the Darwin timescale in V1309~Sco is too long -- likely thousands of years -- 
compared to $\dot P/\ddot P$ inferred from  observations, where $\dot P/\ddot P$ is only about a few years.
Instead, it was proposed that the observed period decay 
is due to non-conservative mass transfer from the primary to the companion accompanied by a simultaneous mass loss via a wind, 
during at least several pre-merger years.

Let us investigate this in more detail.
The tidal friction timescale for a star of mass $M_1$ and radius $R_1$ in a binary with a companion of mass $M_2$ at an orbital separation $a$ can be estimated as \citep{2001ApJ...562.1012E}
\begin{equation}
\tau_{\rm TF} =  \tau_{\rm V} \frac{a^8}{9 R_1^8} \frac{M_1^2}{(M_1+M_2)M_2} (1-Q)^2 \ ,
\label{eq:taurf}
\end{equation}
\noindent where $Q$ is the quadrupolar deformability of the star, $Q=0.223$ for polytropes of $n=3/2$ \citep{2006epbm.book.....E}. 
For our initial binaries, $\tau_{\rm TF}/ \tau_{\rm V}  \approx 40-70$ (the factor varies from 40 to 70 
due to our range of initial conditions and the strong power dependence of the timescale on the ratio $a/R_1$).
Here $\tau_{\rm V}$ is an intrinsic viscous timescale  \citep{1977A&A....57..383Z}.
For a star with a substantial  convective envelope, $\tau_{\rm V}$ is
the timescale on which turbulent friction takes place, or
the global convective turnover timescale. The detailed stellar model from our stellar evolution code gives the global convective turnover time as $\sim260$ days for our primaries, consistent with a simple estimate from the Zahn formula.
Accordingly, $\tau_{\rm TF}$ is 30-50 years.
This value is smaller than the range quoted in  \cite{2013arXiv1307.4088P} by one to two orders of magnitude.
This is due to two main reasons.  First, \cite{2013arXiv1307.4088P} assumes $Q<<1$, which, while appropriate for polytropes with $n=3$, leads to an overestimate of $\tau_{\rm TF}$ by nearly a factor of 2 for stars with large $n=1.5$ convective envelopes.  Second, \cite{2013arXiv1307.4088P} adopts that $\tau_{\rm V}$ can be as large as decades, which is significantly larger than the actual global convective turnover time given by a stellar evolution code or the formula for the friction timescale \citep[Equation (4.11) in][]{1977A&A....57..383Z}.

We further note that originally the Darwin instability and its relation with the viscous timescale
were formally established in the limit of small viscosities (weak friction)  for 
{\it equilibrium tides} \citep{1973Ap&SS..23..459A, 1977A&A....57..383Z, 1981A&A....99..126H,2001ApJ...562.1012E}.
The basic assumption of equilibrium tides is that 
isobaric surfaces within the star are always equipotential surfaces -- i.e., the star
is in the state of hydrostatic equilibrium \citep{2001ApJ...562.1012E}.
It might be expected that this approximation breaks down when a star approaches its Roche lobe.
Indeed, \cite{2012JASS...29..145E} argued that the Darwin instability  in binaries with 
extreme mass ratios and evolved companions
(as in the case of V1309~Sco) can have a timescale as small as a few days.

However, even though the Darwin instability 
has most likely naturally led to the merger both in V1309~Sco and in our simulations, 
it is not likely that it will naturally explain the rate of the orbital decay 
in our simulations.  The artificial viscosity in our simulations can not match completely the convective viscosity that has played a role
prior to merger. 
In a star with a convective envelope, viscosity comes from up and down motions, which are not present in our SPH model.
In the SPH code, an artificial viscosity acts only when a velocity gradient is present, and hence has a very different effect 
on the evolution of a binary orbit than what would be caused by the convective viscosity in a real binary.

We should clarify that the orbital decay due solely to the Darwin instability (before {\it any} mass loss) could have been observed in our simulations only
while $R_{*}$ remains smaller than Roche lobe of the donor-- once $R_{*}$  is larger, the donor would always have some SPH particles
that find themselves outside of its Roche lobe. Those particles may get lost from the binary, may relocate to the Roche lobe of the secondary, 
or may get slightly more energetic and trigger further expansion,
but ultimately these SPH particles oscillations speed up the orbital decay.
In most of our simulations, the significant orbital evolution prior to merger 
takes place when $R_* > R_{\rm RL}$ and is affected by particles oscillations.

Because the timescales cannot be used for a direct comparison, we have to look to other features. 
The simulated {\it shape} of the orbital decay does match the observed exponential shape for
a synchronized binary with a main-sequence companion or for a non-synchronized binary with a degenerate companion.
Based on this pre-merger behavior, these are our favored types of progenitor binary.
In contrast, the orbital decay in synchronized binaries with a degenerate companion is not consistent with the observed exponentially-shaped decay.

Now let us return to the mass loss from the system.
In our simulations, we indeed observe situations in which most of the mass lost from the system prior to merger
proceeds via $L_2$. However, this occurs only in initially synchronized  binaries (note that as synchronized binaries we include systems with the degree of corotation down to 0.85) -- non-synchronized systems 
show mainly an isotropic mass loss in ``clumps.''
Why do synchronized binaries in our simulations have $L_2$ mass loss prior to merger, while non-synchronized do not?
This could be because the specific angular momentum of the material that is transferred from the donor 
to the secondary in a synchronized system is higher
than in an non-synchronized system, and a particle can be lost via $L_2$ only if it has an angular momentum 
high enough to at least reach $L_2$ point.
Indeed, we find that the average specific angular momentum of SPH particles moving in the neighborhood of $L_1$ point toward the secondary 
in synchronized systems exceeds the angular momentum threshold posed by $L_2$ location, but only by about 50\%.
It also might be because only in synchronized systems the donor is shaped as theoretically expected simplified Roche lobe,
while in an non-synchronized binary, the donor is significantly more spherical and starts to lose mass before it extends to $L_1$.

A fully non-conservative $L_2$ mass loss 
in our system leads to a decrease in the specific angular momentum of the remaining binary.
It forces the orbit to shrink, which leads to the increase in Roche lobe overflow;
the latter accelerates exponentially the $L_1$ mass transfer, which  quickly 
becomes fully dynamical, and the system merges.
Indeed, in our simulations, $L_2$ mass loss, once started, lasts only for a few days.
However, non-synchronized binaries, those that do not have $L_2$ mass loss, merge too, even though 
it takes much longer for them to complete the merger after their $L_1$ mass transfer started.
Hence, depending on how well the system is synchronized, $L_2$ mass loss can precede the merger, 
but it does not have to be responsible for the merger to occur.
Our results therefore suggest that $L_2$ mass loss could not be responsible for 
a long-term (several years timescale) orbital decay in V1309~Sco
\footnote{We also find that  
Equation (3) in \cite{2013arXiv1307.4088P}  is incorrect, as can be verified by checking this equation in the limiting case of $\beta=1$, when the mass is lost with specific angular momentum of the donor star.  
 \cite{2013arXiv1307.4088P} uses equation (3) for the orbital period evolution during the mass transfer, and at the same
time for calculating mass transfer rates, and, consequently, for the mass loss rate. Detailed derivations of orbital evolution for various modes of the mass loss and the mass transfer can be found in \cite{1997A&A...327..620S}.}.
This, coupled with our estimate for $\tau_{\rm TF}$ above and with our checks of the Darwin instability criterion,
advocates that it was indeed the Darwin instability that resulted in the observed orbital decay.

\cite{2012ApJ...746..100S} discussed that the outbursts for ILOTs could be powered by mass accretion 
onto a main-sequence star which could potentially launch jets. We found that the non-synchronized cases do not form an accretion disk, since the donor's particles do not have enough angular momentum to go through $L_1$. \revone{In contrast, the synchronized cases show a few SPH particles around the main-sequence star (accretion disk). The total mass of these particles is $\sim0.0005M_\odot$, and their velocities, relative to the main-sequence-star centre of mass, are up to 200 km/s.  For comparison, the escape velocity of the main sequence star is close to 600 km/s.} The highest velocity gas in our simulations which is just above the escape velocity of the donor is reached when the CE is starting or has started; hence, the highest velocity gas is not due to jets/winds from the accretion disk. The accompanying visualizations are useful for understanding the flow pattern of the gas (see the on-line-only visualizations \ref{mov:num1} and \ref{mov:num2}).

We have analyzed how the mass loss proceeded throughout all of our simulations, finding that most of the mass loss takes place 
in up to 3 individual mass outburst -- before the merger, during the merger, and after the merger was completed --  
where each outburst takes away from about $0.0048$ to $0.047$ $M_\odot$ and
lasts from one to a few days (several dynamical timescales of the initial binary).
Our synchronized systems with a degenerate companion 
lack a clear separation between the mass outbursts before and during the merger. 
All our simulations with a non-degenerate donor have two episodes 
of mass outburst and lack the third mass outburst after the merger is completed.
All simulations with a non-synchronized donor and a degenerate companion show 3 mass outbursts, and vice versa.
The observed light-curve was reconstructed best with 2 mass outbursts, suggesting that the latter 
systems are least likely to represent the initial binary. 
The total amount of the ejected mass in our simulations does not vary much between the models and is from $0.038 M_\odot$ to $0.086 M_\odot$.

The kinetic energy of the ejected material at infinity is comparable to the initial binding energy in the envelope of the donor. 
This suggests that the energy formalism used for predicting common envelope outcomes needs to account for kinetic 
energies in the energy budget.
We find that velocities at infinity are 160-190~km~s$^{-1}$; they are in the same range as were found in observations of V1309~Sco using the profile of 
the HeI emission line, 160-180~km~s$^{-1}$  \citep{2010A&A...516A.108M}.

We find that the specific angular momentum of the ejected material is significantly larger than the specific angular momentum of the material
in the initial binary, by 5.8 to 8.1 times\footnote{We note that in the $\gamma$-formalism that uses angular momentum conservation to predict
the outcomes of the CE events, this value would be taken as 1.5 \citep{2000A&A...360.1011N}. There is no reason to believe that the same 
value of $\gamma \sim 1.5$ be valid for all the systems entering the common envelope phase.
Since this parameter has a tremendous effect on the outcome, where a difference in only few per cent in its value can change 
the outcome completely \citep{2011ASPC..447..127W, 2013A&ARv..21...59I},  a fine-tuning for each kind of binary needs to be done.}.
In our simulations involving a synchronized binary, the ejecta form a well distinguished outer ring   with a bit less pronounced second inner ring --
this is the consequence of $L_2$ mass loss.
In non-synchronized binaries, the ejecta have the shape of an expanding bubble with ``clumps.''
Currently, the observations do not yet allow us to distinguish whether the ejecta form a ring or bubble \citep{2013MNRAS.431L..33N},
but it may be done in the future. In this case, it can provide a further insight on how much the system was synchronized at the start.

\cite{2011A&A...535A..50M} proposed that if a merger produces a disk, this disk could be the 
progenitor of Jupiter-like planets around the merged star. We observe that ejecta is ``clumped'' in all our simulations, with the clumpiness being especially apparent in the case of non-synchronized systems. 
The long-term evolution of these clumps can not be traced in our code.

After the merger, hydrostatic equilibrium in the bound mass is obtained fairly quickly, within a dozen days.
By then, most of the bound and optically thick mass is located within a radius of 100 $R_\odot$. 
This luminous and expanded object will further experience thermal relaxation, with an initial timescale of a few years.
We find that the formed star is significantly shock heated compared to its progenitor and, before thermal relaxation takes place,
has an entropy profile characteristic of a radiative star. With our SPH code, 
we can not judge when exactly the envelope of the star will become convective again, but it may take place as quickly as within a few years after the merger.

Partially as a result of the high angular momentum loss with the ejecta, and partially due to a relatively slow rotation 
of the initial system, the merged objects are not expected to be necessarily at their critical rotation even 
after they are thermally relaxed and shrunk -- their angular momenta are only about 2-3 times of the initial ones.
The merged star, when in hydrostatic equilibrium, but before its thermal equilibrium, 
does not appear to have a solid body rotation -- while the donor is still expanded, 
its outer layers rotate significantly slower than its inner layers. 
For a time after the thermal relaxation (which takes only a few years for the expanded outer layers),
the obtained specific angular momentum profile predicts that the outer layers will rotate faster than the inner layers.
It may be expected that during the thermal relaxation, 
as the object transforms from a radiative to a convective star, and at the same time would
attempt to redistribute the angular momentum, the object will have strong differential rotation.
This may result in an efficient dynamo operation that will be accompanied by X-ray luminosity \citep{2007MNRAS.375..909S}.
Following the derivation in \cite{2007MNRAS.375..909S}, this X-ray luminosity 
can be estimated to be $\sim 10^{31}$ erg s$^{-1}$ during the envelope 
contraction phase.  We note however that their estimate might be not fully applicable, 
as at the initial contraction stage the object is radiative.  A more detailed study of how the thermal relaxation proceeds in the merger product 
is definitely required for understanding magnetic field formation when the convective envelope is developed for the first time.
Currently, 5 years after the outburst, 
 the ejecta provides a hydrogen column density from $\sim 10^{23}$ to $ 10^{24}$ cm$^{-2}$ 
and can hide an X-ray object with the luminosity up to $10^{32}$ erg s$^{-1}$.
Indeed, a recent Chandra observation, made in 2013,  
did not detect a single photon during 35 ksec exposure (S.~Rappaport, private communication).

We conclude that all considered progenitor binaries can produce an outburst resembling the V1309~Sco event.
The comparison of details of observations with such features obtained in simulations as how the mass is ejected, 
what is the radius of the merged object, and the shape of the orbital period decay before the merger 
favors most a synchronized binary with a main sequence companion.
Future observations of the shape of the ejected material (a shell-type bubble versus a ring) 
and the X-ray luminosity can help with further understanding of V1309 Sco object.


\acknowledgments
The authors are grateful to  S.~Rappaport for providing the results of Chandra observations. The authors thank R. Tylenda and the referee S. Noam for the valuable comments.
J.L.A.N. acknowledges support from CONACyT.  N.I. acknowledges support by NSERC Discovery Grants and Canada Research Chairs Program; this research was supported in part by the National Science Foundation under Grant No. NSF PHY05-51164.
JCL is supported by the National Science Foundation under grant No. AST-1313091 and thanks Zach Silberman for useful discussions.  Some code for this project was developed in the Extreme Science and Engineering Discovery Environment (XSEDE), which is supported by National Science Foundation grant number OCI-1053575. This research has been enabled by the use of computing resources provided by WestGrid and Compute/Calcul Canada.

\appendix

\section{Entropy in a Fully Ionized Gas}
\label{app:a}

The specific entropy $s$ (erg K$^{-1}$ g$^{-1}$) of a mixture of atoms, ions, and electrons together with radiation is given by  \cite{2001spfc.book.....B} as follows,

\begin{equation}
\label{eq:totalentropy}
s=\frac{k}{\rho}\sum_{i}\sum_{j=0}^{i}n_{ij}\left\{ \frac{5}{2}+\ln\left[\left(\frac{m_ikT}{2\pi \hbar^2}\right)^{3/2}\frac{g_{ij}}{n_{ij}}\right]\right\} 
+\frac{k}{\rho}n_e\left\{\frac{5}{2}+\ln\left[\left(\frac{m_ekT}{2\pi \hbar^2}\right)^{3/2}\frac{2}{n_e}\right]\right\}+\frac{4}{3} \frac{aT^3}{\rho},
\end{equation}
where $x_i$ is the mass fraction of the element, $y_{ij}$ is the fraction of the $i$-th element ionized to the $j$-state, $m_i=A_im_{\rm u}=(1{\rm g})A_i/N_A$ is the nuclear mass,
\begin{equation}
 n_{ij}=x_i\rho y_{ij}/m_i
\label{eq:numberdensity}
\end{equation}
is the number density of ions,
\begin{equation}
 n_e=\sum_i\sum_{j=1}^i j n_{ij} \mbox{ cm}^{-3}
\end{equation}
is the electron number density,
\begin{equation}
 \mu=\left[\sum_i \frac{m_u}{m_i}\ x_i\sum_{j=0}^i \left(1+j\right)y_{ij}\right]^{-1}
\end{equation}
is the number of nucleons, $m_{\rm u}=1.66057\times 10^{-24}$ g is the atomic mass unit, $N_A=(1\mbox{ g})/m_{\rm u}\ \mbox{ mole}^{-1}$ is the Avogadro's number, $m_e=9.10953\times10^{-28}$ g is the mass of 
the electron, $\hbar=1.0546\times10^{-27}$ erg s is the Planck constant, $k=1.38064\times10^{-16}$ erg K$^{-1}$ is the Boltzmann constant,
$a=7.565\times10^{-15}$ erg cm$^{-3}$ K$^{-4}$ is the radiation density constant and $c=2.9979\times 10^{10}$ cm s$^{-1}$ is the velocity of light in vacuum.
 For a fully ionized gas the fraction of the $i$-th element ionized to the $j$-state can be written as
\begin{equation}
 y_{ij}=\left\{
\begin{array}{cc}
 1&i=j\\
 0&i\neq j
\end{array}
\right.,
\end{equation}
and $A_i\approx 2i$. Then, we can get 
\begin{equation}
 n_e=\sum_i i n_{ii}=\rho\sum_i i \frac{x_i}{m_i} =\frac{\rho N_A}{\mu_e},
\label{eq:electrondensityb}
\end{equation}
where $\mu_e$ is the mean molecular weight per free electron,
$$\mu_e\equiv\left[\sum_i i \frac{x_i}{A_i}\right]^{-1}\approx\left[x_H+\frac{x_{He}}{2}+\frac{1}{2}x_A\right]^{-1};\ \ \ \ \ x_A=\sum_{i\geq 3}x_i.$$
For the case of ions, we can write the total density as follows,
\begin{equation}
 n_I=\rho\sum_i \frac{x_i}{m_i}=\frac{\rho N_A}{\mu_I},
\label{eq:ions}
\end{equation}
where $\mu_I$ is the ion mean molecular weight,
$$\mu_I\equiv\left[\sum_i \frac{x_i}{A_i}\right]^{-1}\approx\left[x_H+\frac{x_{He}}{4}+\frac{1}{14}x_A\right]^{-1};\ \ \ \ \ x_A=\sum_{i\geq 3}x_i.$$
Notice that we assume that the average of $A_i$ is about 14. Now, we can define $X=x_{\rm H}$, $Y=x_{\rm He}$ and $Z=\sum x_i$ as the elements heavier than Helium. Therefore,
\begin{equation}
 X+Y+Z=1,
\end{equation}
and we can re-write $\mu_I$ and $\mu_e$ as follows,
\begin{eqnarray}
 \mu_I&\approx&\frac{28}{26X+5Y+2},\\
 \mu_e&\approx&\frac{2}{X+1}.
\end{eqnarray}
Hence, 
\begin{equation}
 \mu=\left(\frac{1}{\mu_I}+\frac{1}{\mu_e}\right)^{-1}=\frac{28}{40X+5Y+16}.
\end{equation}
Hence, the entropy of the gas can be written as follows, by substituting the equations \eqref{eq:numberdensity} and \eqref{eq:electrondensityb} into \eqref{eq:totalentropy} and expanding terms,
\begin{eqnarray*}
 s&=& kN_A\sum_{i}\frac{x_i}{A_i}\ln\left[\left(\frac{kT}{N_A2\pi \hbar^2}\right)^{3/2}\frac{1}{\rho N_A}\right]+ kN_A\sum_{i}\frac{x_i}{A_i}\ln\left(\frac{A_i^{5/2}}{x_i}\right) +\frac{5}{2} \frac{kN_A}{\mu_I}+\frac{5}{2} \frac{kN_A}{\mu_e} +\frac{4}{3} \frac{aT^3}{\rho} \\
&+&\frac{kN_A}{\mu_e}\ln\left[\left(\frac{kT}{N_A2\pi \hbar^2}\right)^{3/2}\frac{1}{\rho N_A}\right]+\frac{kN_A}{\mu_e}\ln\left[2 (m_eN_A)^{3/2}\mu_e\right],
\end{eqnarray*}
which has terms in common, thus, the new equation can be written as
\begin{eqnarray*}
 s&=& \frac{kN_A}{\mu}\ln\left[\left(\frac{kT}{N_A2\pi \hbar^2}\right)^{3/2}\frac{1}{\rho N_A}\right]
+\frac{5}{2} \frac{kN_A}{\mu} +\frac{4}{3} \frac{aT^3}{\rho} +\frac{kN_A}{\mu_e}\ln\left[2 (m_eN_A)^{3/2}\mu_e\right]
+ kN_A\sum_{i}\frac{x_i}{A_i}\ln\left(\frac{A_i^{5/2}}{x_i}\right). 
\end{eqnarray*}
Hence, we can rearrange the previous equation and get
\begin{equation}
 s=\frac{kN_A}{\mu}\ln\left(T^{3/2}\rho^{-1}\right)+\frac{4}{3} \frac{aT^3}{\rho}+s_{0},
 \label{eq:specentropyfully}
\end{equation}
where
\begin{eqnarray*}
 s_{0}&=&\frac{kN_A}{\mu}\left[\frac{5}{2}+\ln\left(\frac{k}{2\pi N_A^{5/3} \hbar^2}\right)^{3/2}\right] +\frac{kN_A}{\mu_e}\ln\left[2 (m_eN_A)^{3/2}\mu_e\right]+kN_A\sum_{i}\frac{x_i}{A_i}\ln\left(\frac{A_i^{5/2}}{x_i}\right).
\end{eqnarray*}
Therefore, the entropy of the $i$-th particle can be written as follows,
\begin{equation}
 S_i=\frac{kN_Am_i}{\mu}\ln\left(T^{3/2}\rho^{-1}\right)+\frac{4}{3} \frac{m_iaT^3}{\rho}+S_{0,i},
 \label{eq:totalentropyfully}
\end{equation}
where 
\begin{equation}
 S_{0,i}\equiv m_is_{0,i} =\frac{kN_Am_i}{\mu}\left[\frac{5}{2}+\ln\left(\frac{k}{2\pi N_A^{5/3}\hbar^2}\right)^{3/2}\right] +\frac{kN_Am_i}{\mu_e}\ln\left[2 (m_eN_A)^{3/2}\mu_e\right]+kN_Am_i\sum_{j}\frac{x_j}{A_j}\ln\left(\frac{A_j^{5/2}}{x_j}\right).
\label{eq:entropyconstcorr}
\end{equation}
Here we used $g_{ii}=1$.
 J.L.A


\bibliographystyle{apj}
\bibliography{references}


\begin{figure}[htp!]
  \begin{center}
   \includegraphics[scale=0.35]{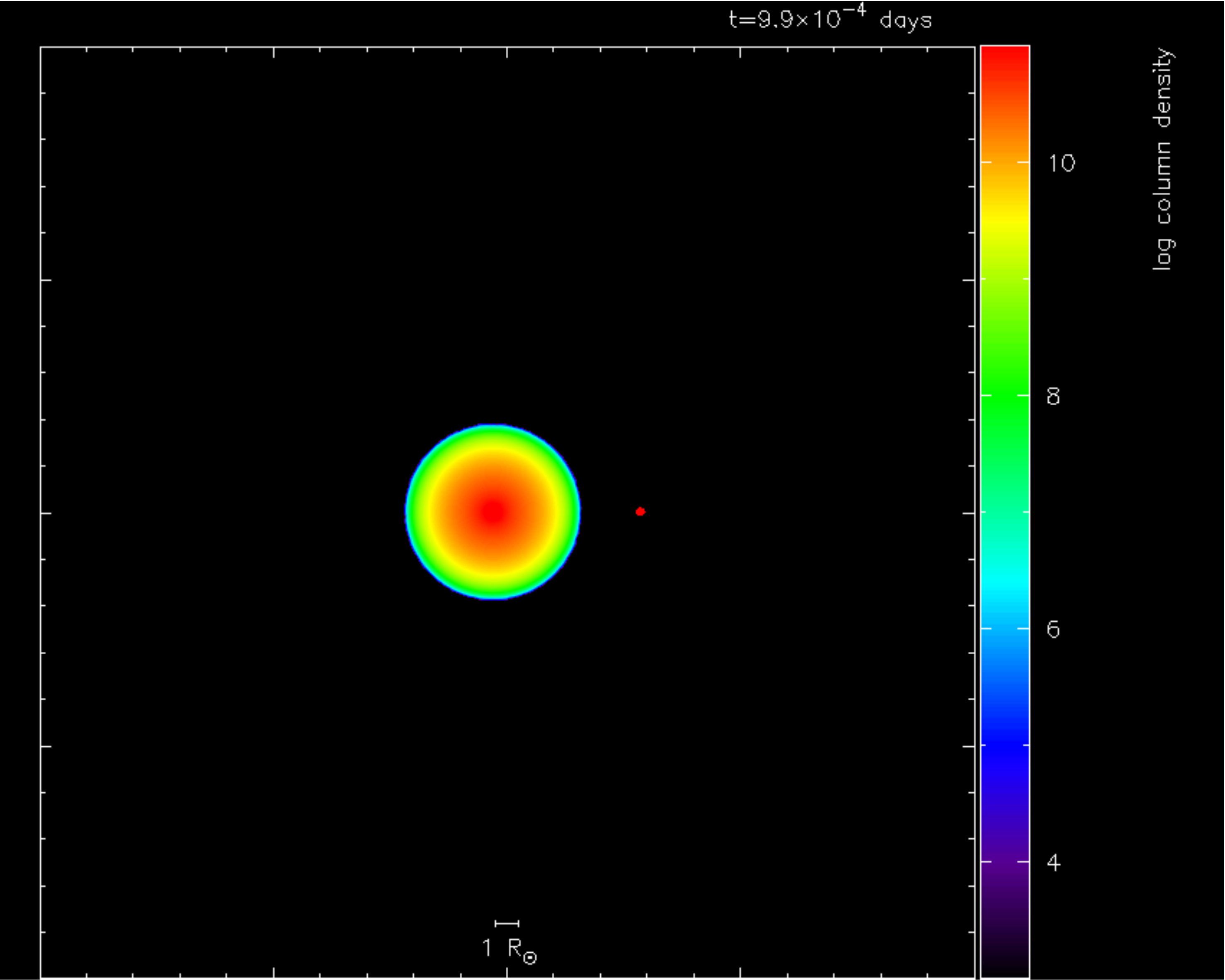}
 \end{center}
 \caption{Column density animation showing the merger of a non-synchronized binary containing a $M=1.52M_\odot$, $R=3.51R_\odot$ giant and a $M=0.16M_\odot$ degenerate companion with an initial orbital period of $1.42$ days (simulation pn351). This animation is available only online.}
\label{mov:num1}
\end{figure}

\begin{figure}[htp!]
 \begin{center}
   \includegraphics[scale=0.35]{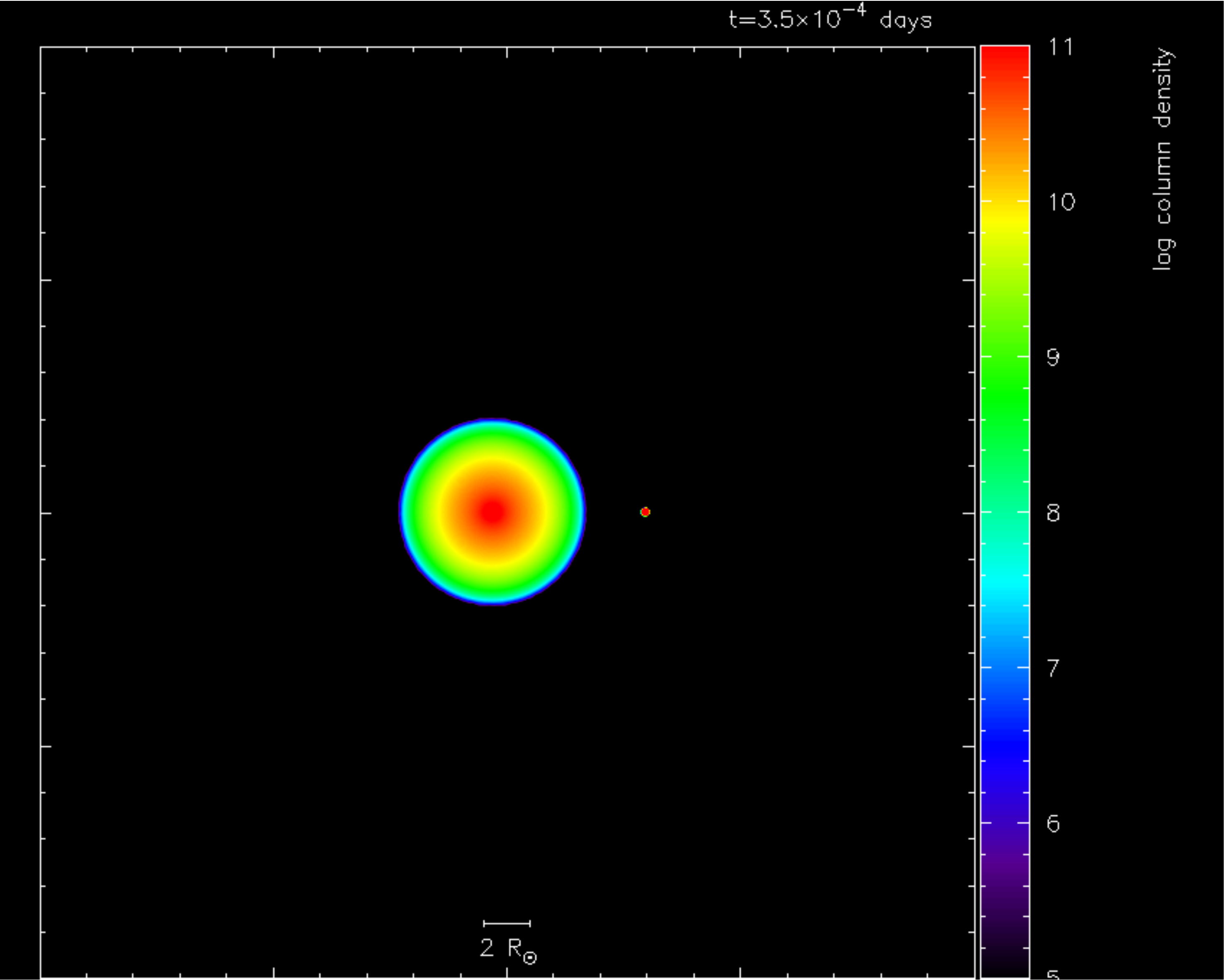}
 \end{center}
 \caption{Column density animation showing the merger of a synchronized binary containing a $M=1.52M_\odot$, $R=3.76R_\odot$ giant and a $M=0.16M_\odot$ main-sequence companion with an initial orbital period of $1.50$ days (simulation ms376). This animation is available only online.}
\label{mov:num2}
\end{figure}


\end{document}